\documentclass[accepted=2023-04-10,a4paper,onecolumn]{quantumarticle}
\pdfoutput=1
\PassOptionsToPackage{hyphens}{url}
\usepackage[pagebackref=true,colorlinks]{hyperref}
\usepackage{amsmath,amssymb,amsthm,mathtools,mathrsfs}
\usepackage{pifont}
\usepackage{bbm}
\usepackage{tikz}
\usetikzlibrary{calc}
\usepackage{adjustbox}
\usepackage{stmaryrd}
\usepackage{fancyhdr}
\usepackage{dsfont}
\usepackage[normalem]{ulem}
\usepackage[bottom]{footmisc}
\usepackage{enumerate}

\usepackage{tocbasic}
\DeclareTOCStyleEntry[
  beforeskip=0.3em plus 1pt,
  pagenumberformat=\textbf
]{tocline}{section}

\makeatletter
\renewcommand*\env@matrix[1][\arraystretch]{%
  \edef\arraystretch{#1}%
  \hskip -\arraycolsep
  \let\@ifnextchar\new@ifnextchar
  \array{*\c@MaxMatrixCols c}}
\makeatother

\theoremstyle{plain}
\newtheorem{theorem}[equation]{Theorem}
\newtheorem{lemma}[equation]{Lemma}
\newtheorem{proposition}[equation]{Proposition}
\newtheorem{corollary}[equation]{Corollary}
\theoremstyle{definition}
\newtheorem{definition}[equation]{Definition}
\newtheorem{construction}[equation]{Construction}
\newtheorem{question}[equation]{Question}

\newtheorem{problem}[equation]{Problem}
\newtheorem{example}[equation]{Example}
\newtheorem{exercise}[equation]{Exercise}
\newtheorem*{answer}{Answer}
\newtheorem*{solution}{Solution}
\newtheorem{remark}[equation]{Remark}

\newtheorem{notation}[equation]{Notation}
\newtheorem{noterm}[equation]{Notation and Terminology}

\newcommand\define[1]{\emph{\textbf{#1}}}

\numberwithin{equation}{section}

\let\C=\Chi

\newcommand{\be}{\begin{equation}}
\newcommand{\ee}{\end{equation}}
\def\ba{\begin{align}} 
\def\ea{\end{align}}
\newcommand{\bea}{\begin{eqnarray}}
\newcommand{\eea}{\end{eqnarray}}
\newcommand{\bx}{\begin{example}}
\newcommand{\ex}{\end{example}}
\newcommand{\bex}{\begin{exercise}}
\newcommand{\eex}{\end{exercise}}
\newcommand{\ban}{\begin{answer}}
\newcommand{\ean}{\end{answer}}
\newcommand{\bt}{\begin{theorem}}
\newcommand{\et}{\end{theorem}}
\newcommand{\bc}{\begin{corollary}}
\newcommand{\ec}{\end{corollary}}
\newcommand{\blem}{\begin{lemma}}
\newcommand{\elem}{\end{lemma}}
\newcommand{\bp}{\begin{problem}}
\newcommand{\ep}{\end{problem}}
\newcommand{\bn}{\begin{proposition}}
\newcommand{\en}{\end{proposition}}
\newcommand{\bd}{\begin{definition}}
\newcommand{\ed}{\end{definition}}
\newcommand{\bcon}{\begin{construction}}
\newcommand{\econ}{\end{construction}}
\newcommand{\bq}{\begin{question}}
\newcommand{\eq}{\end{question}}
\newcommand{\bprf}{\begin{proof}}
\newcommand{\eprf}{\end{proof}}
\newcommand{\br}{\begin{remark}}
\newcommand{\er}{\end{remark}}
\newcommand{\bs}{\begin{solution}}
\newcommand{\es}{\end{solution}}
\newcommand{\beqs}{\begin{eqnarray}}
\newcommand{\eeqs}{\end{eqnarray}}
\newcommand{\bnt}{\begin{noterm}}
\newcommand{\ent}{\end{noterm}}
\newcommand{\bnot}{\begin{notation}}
\newcommand{\enot}{\end{notation}}

\newcommand{\<}{\langle}
\renewcommand{\>}{\rangle}

\def\td{\tilde}

\newcommand{\id}{\mathrm{id}}

\newcommand{\tr}{{\rm tr} }

\def\R{{{\mathbb R}}}
\def\C{{{\mathbb C}}}

\def\Z{{{\mathbb Z}}}

\newcommand{\Ad}{\mathrm{Ad}}

\DeclareMathAlphabet{\mathpzc}{OT1}{pzc}{m}{it}
\newcommand{\Alg}[1]{\mathpzc{#1}}

\def\mE{{{\mathcal{E}}}}
\def\mF{{{\mathcal{F}}}}

\newcommand{\matr}{\mathbb{M}}

\newcommand{\op}{\mathrm{op}}
\newcommand{\St}{\mathbf{States}}
\newcommand{\CSt}{\mathbf{CStates}}

\newcommand{\CPTP}{\mathbf{CPTP}}

\newcommand{\shriek}{\operatorname{!}}

\def\retro{\mathscr{R}}
\def\Petz{\mathrm{P}}
\def\STH{\mathrm{STH}}
\def\JRSWW{\mathrm{JRSWW}}
\def\SS{\mathrm{SS}}
\def\Bayes{\mathrm{B}}

\newcommand{\ben}{\renewcommand{\theenumi}{\alph{enumi}} 
\renewcommand{\labelenumi}{(\theenumi)}\begin{enumerate}}
\newcommand{\een}{\end{enumerate}}
\newcommand{\cmark}{\ding{51}}
\newcommand{\xmark}{\ding{55}}%

\newcommand\blfootnote[1]{%
  \begingroup
  \renewcommand\thefootnote{}\footnote{#1}%
  \addtocounter{footnote}{-1}%
  \endgroup
}

\title{Axioms for retrodiction: 
achieving time-reversal symmetry with a prior}
\author{Arthur J.~Parzygnat and Francesco Buscemi}
\date{2023-05-12}

\newcommand{\Addresses}{{
  \bigskip
  \footnotesize
  
    A.~J.~Parzygnat, \textsc{
  Graduate School of Informatics, Nagoya University, Chikusa-ku, 464-8601 Nagoya, Japan}\par\nopagebreak
  \textit{E-mail address}, A.~J.~Parzygnat: \texttt{parzygnat@nagoya-u.jp}, \texttt{parzygnat@ihes.fr}

  \medskip

    F.~Buscemi, \textsc{Graduate School of Informatics, Nagoya University, Chikusa-ku, 464-8601 Nagoya, Japan}\par\nopagebreak
  \textit{E-mail address}, F.~Buscemi: \texttt{buscemi@i.nagoya-u.ac.jp}

}}

\begin{document}
\emergencystretch 2em

\maketitle
\begin{abstract}  
We propose a category-theoretic definition of retrodiction and use it to exhibit a time-reversal symmetry for all quantum channels. 
We do this by introducing retrodiction families and functors, which capture many intuitive properties that retrodiction should satisfy and are general enough to encompass both classical and quantum theories alike. 
Classical Bayesian inversion and all rotated and averaged Petz recovery maps define retrodiction families in our sense. 
However, averaged rotated Petz recovery maps, including the universal recovery map of Junge--Renner--Sutter--Wilde--Winter, do not define retrodiction functors, since they fail to satisfy some compositionality properties. 
Among all the examples we found of retrodiction families, the original Petz recovery map is the only one that defines a retrodiction functor.
In addition, retrodiction functors exhibit an inferential time-reversal symmetry consistent with the standard formulation of quantum theory.
The existence of such a retrodiction functor seems to be in stark contrast to the many no-go results on time-reversal symmetry for quantum channels. 
One of the main reasons is because such works defined time-reversal symmetry on the category of quantum channels alone, whereas we define it on the category of 
quantum channels \textit{and} quantum states.
This fact further illustrates the importance of a prior in time-reversal symmetry.


\end{abstract}

\vspace{-7mm}
\tableofcontents

\section{Retrodiction versus time reversal}

As we currently understand them, the laws of physics for closed systems are time-reversal symmetric in both classical and quantum theory.  
The associated evolution is reversible in the sense that if one evolves any state to some time in the future, 
one can (in theory) apply the reverse evolution, which is unambiguously defined, to return the initial state. 

However, not all systems of interest are closed, and so, not all evolutions are reversible in the sense described above. For instance, arbitrary stochastic maps and quantum channels are of this kind. This, together with other phenomena such as the irreversible change due to a measurement~\cite{vN18,Lu06,Kr83} and the black hole information paradox~\cite{Ha75,Ha76,Ha82}, has led many to question the nature of time-reversibility. Indeed, many have provided no-go theorems and occasionally offered proposals for what time-reversal is or how it can be implemented (physically or by belief propagation)~\cite{Wat55,ABL64,BPJ00,BaKn02,Cr08,APT10,LeSp13,Ba14,AZZ15,OrCe15,OrCe16,Le16,CGS17,LePu17,Oe19,BJP21,FSB20,DDR21,CAZ21,Ha21,ChLi22,SSGC22}. The following lists some of the questions considered in these programs.

\begin{itemize}
\item
In what sense can channels be reversed to construct a time-symmetric formulation of open dynamics?
Namely, what axioms \emph{define} what we mean by time-reversibility~\cite{CAZ21,ChLi22,SSGC22}?
\item
As opposed to applying classical probabilistic inference associated with experimental outcomes~\cite{Le07,Le16,LePu17,DDR21}, is it possible to provide a fully \emph{quantum} formulation of inference that does not require the classical agent interface?
\item
Is the apparent directionality of time a consequence of the irreversibility of certain processes that are more general than deterministic evolution? 
\item
What is a maximal subset of quantum operations that has a reasonable time-reversibility? For example, it is known that a certain type of reversibility is possible for unital quantum channels~\cite{CAZ21,DDR21,ChLi22,CGS17,SSGC22} and more generally for quantum channels that fix an equilibrium state~\cite{Cr08}. Is this the best we can do without modifying quantum theory~\cite{OrCe15}?
\end{itemize}

Although a large body of work has recently focused on quantum theory, a similar problem lies in the classical theory. More importantly, it has been understood for a long time that complete reversibility is much too stringent of a constraint to ask for. Instead, one might be more interested in \emph{retrodictability} rather than \emph{reversibility}~\cite{Wat55,BuSc21}, where retrodictability is the ability to \emph{infer} about the past and which specializes to reversibility in the case of reversible dynamics.
And in the classical theory, retrodictability is achieved through Bayes' rule and Jeffrey's probability kinematics, the latter of which specializes to Bayes' rule in the case of definitive evidence, but is strictly more general in that an arbitrary state can be used as evidence to make inference~\cite{Ba1763,Pearl88,Je90,Ba14}.

But while the same Bayes' rule and Jeffrey's probability kinematics can be used simultaneously for both spatial and temporal correlations in classical statistics, quantum theory seems to reveal a subtle distinction between these two forms of inference~\cite{LeSp13,HHPBS17,FuPa22,FuPa22QB}. The study of the unique spatial correlations in quantum theory has been a subject of intense studies since at least the work of Schr\"odinger~\cite{EPR,Sc35,Sc36,WJD07}. Meanwhile, our understanding of the temporal correlations unique to quantum theory is not as well-understood and is still under  investigation~\cite{LeGa85,FJV15,FuPa22QB}. 
In order to contribute towards this important and evolving subject, it is retrodictability, i.e., the inferential form of time-reversal symmetry, in quantum (and classical) dynamics that will be the focus of this paper.

However, retrodictability is not in general possible for arbitrary dynamics unless additional input data are provided. One example of the kind of information that can be used is a \emph{prior}, which summarizes the state of knowledge of the retrodictor. Previous proposals that have avoided the introduction of a prior oftentimes secretly included it. For example, unital quantum channels are precisely quantum channels that preserve the uniform prior, even though this is not always phrased in this way because the identity matrix disappears from expressions.

In the present paper, we show that a form of time-reversal symmetry is possible for certain open system evolutions (classical and quantum channels) \emph{together with} priors, i.e., we prove the existence of an inferential time-reversal symmetry for open quantum system dynamics.
This bypasses some of the no-go theorems in the literature, such as in~\cite{CGS17,CAZ21,ChLi22}, in at least two ways. First, by including the prior, the assignment of a time-reversal no longer has input just some channel, but a channel and a prior. 
Second, we see no reason to assume that the time-reversal assignment must be linear in either the channel or the prior, as is done in~\cite{CAZ21,ChLi22} (see axiom~4 on page~2 in~\cite{ChLi22} or the definition of symmetry in~\cite{CAZ21}). Indeed, even classical Bayesian inversion is linear in neither the original process nor the prior, so there is no reason to assume linearity in the quantum setting, which should specialize to the classical theory. 
In fact, in the present work, we axiomatize retrodiction in a way that is agnostic to classical or quantum theory, probability theory, and even any physical theory as long as it has sufficient structure to describe states and evolution. This is achieved by formulating a rigorous definition of retrodiction in the language of category theory, the proper mathematical language for describing processes~\cite{Se07,CoKi17,ChJa18,Fr20,PaBayes,SSGC22}. 

Fortunately, it is possible to summarize our axioms intuitively without requiring familiarity with category theory.%
\footnote{In the body of this work, only elementary ideas from category theory will be used to formalize our definitions and results. If the reader is comfortable with the definitions of a monoidal category, its opposite, and functor, this should suffice. 
We will try to ease into this through our setting of retrodiction rather than giving formal definitions. The reader is encouraged to read~\cite{Da17} for a first impression of category theory and then browse the reference~\cite[Chapter~1]{Pe19} for a more detailed, yet friendly, introduction. References that cover advanced topics while maintaining a close connection to quantum physics include~\cite{BaSt11,CoKi17,HeVi19}. Finally, standard in-depth references include~\cite{Ri17,Ma98}. 
}
 In order to be agnostic to classical or quantum theory, we try to avoid using language that is only applicable to either. To describe retrodiction, we should first specify the collection of allowed systems (denoted $\Alg{A},\Alg{B},\Alg{C},\dots$), processes (denoted $\mE,\mF,\dots$), and states (denoted $\alpha,\beta,\gamma,\dots$). Having done this, and as argued above, in order to define the retrodiction of some process $\mathcal{E}:\Alg{A}\to\Alg{B}$ from one system $\Alg{A}$ to another $\Alg{B}$, it is necessary to have some additional input. For us, that input is a prior $\alpha$, which is a state on system $\Alg{A}$. Hence, we will combine these data together by writing $(\alpha,\mathcal{E})$. Retrodiction should be some assignment whose input is such a pair $(\alpha,\mathcal{E})$ and whose output is some map $\retro_{\alpha,\mathcal{E}}:\Alg{B}\to\Alg{A}$ that only depends on this input (this is sometimes called \emph{universality} in the literature). 
 
The definition of retrodiction that we propose consists of the following \emph{logical} axioms on such an assignment $\retro$ over all systems $\Alg{A},\Alg{B},\Alg{C},\dots$, evolutions (i.e., processes) $\mE,\mF,\dots$, and states $\alpha,\beta,\gamma,\dots$ (remarks and clarifications on these axioms follow immediately after).
\begin{enumerate}
\item
Retrodiction should produce valid processes, i.e., 
the map $\retro_{\alpha,\mathcal{E}}$ should be a valid process.
\item
The map $\retro_{\alpha,\mathcal{E}}$ should take the \emph{prediction} $\mathcal{E}(\alpha)$ back to the prior $\alpha$, i.e., $\retro_{\alpha,\mathcal{E}}(\mathcal{E}(\alpha))=\alpha$.
\item
The retrodiction of the process that does nothing is also the process that does nothing, i.e., $\retro_{\alpha,\id_{\Alg{A}}}=\id_{\Alg{A}}$.
\item
More generally, the retrodiction of a genuinely reversible process
$\mathcal{E}:\Alg{A}\to\Alg{B}$ is the inverse $\mathcal{E}^{-1}$ of the original process, i.e., $\retro_{\alpha,\mathcal{E}}=\mathcal{E}^{-1}$, and is therefore independent of the prior. 
\item
Retrodiction is involutive in the sense that retrodicting on a retrodiction gives back the original process, i.e., $\retro_{\mathcal{E}(\alpha),\retro_{\alpha,\mathcal{E}}}=\mathcal{E}$.
\item
Retrodiction is compositional
 in the sense that if one has a prior $\alpha$ on system $\Alg{A}$ and two successive processes $\mathcal{E}:\Alg{A}\to\Alg{B}$ and $\mathcal{F}:\Alg{B}\to\Alg{C}$, then the retrodiction $\retro_{\alpha,\mathcal{F}\circ\mathcal{E}}$ associated with the composite process $\Alg{A}\xrightarrow{\mathcal{E}}\Alg{B}\xrightarrow{\mathcal{F}}\Alg{C}$ is the composite of the retrodictions $\retro_{\mathcal{E}(\alpha),\mathcal{F}}:\Alg{C}\to\Alg{B}$ and $\retro_{\alpha,\mathcal{E}}:\Alg{B}\to\Alg{A}$ associated with the constituent components, i.e., $\retro_{\alpha,\mathcal{F}\circ\mathcal{E}}=\retro_{\alpha,\mathcal{E}}\circ\retro_{\mathcal{E}(\alpha),\mathcal{F}}$. 
\item
Retrodiction is tensorial
 in the sense that if one has priors $\alpha$ and $\alpha'$ on systems $\Alg{A}$ and $\Alg{A}'$, respectively, as well as two processes $\mathcal{E}:\Alg{A}\to\Alg{B}$ and $\mathcal{E}':\Alg{A}'\to\Alg{B}'$, then the retrodiction $\retro_{\alpha\otimes\alpha',\mathcal{E}\otimes\mathcal{E}'}:\Alg{B}\otimes\Alg{B}'\to\Alg{A}\otimes\Alg{A}'$ associated with the tensor product of the systems and processes is equal to the tensor product $\retro_{\alpha,\mathcal{E}}\otimes\retro_{\alpha',\mathcal{E}'}$ of the constituent retrodictions, i.e., $\retro_{\alpha\otimes\alpha',\mathcal{E}\otimes\mathcal{E}'}=\retro_{\alpha,\mathcal{E}}\otimes\retro_{\alpha',\mathcal{E}'}$. 
\end{enumerate}

Several clarifications and remarks are in order with regard to these axioms, which we enumerate in the same order. 

\begin{enumerate}
\item
Part of our axioms assume that we have identified a class of physical systems and processes.
In the setting of quantum theory, and in this work, physical systems $\Alg{A},\Alg{B},\Alg{C},\dots$ will be mathematically represented by their corresponding algebras, following the conventions of prior works on quantum inference~\cite{LeSp13}. 
We will take processes to be represented by quantum channels, i.e., completely positive trace-preserving (CPTP) maps (traces on such algebras are defined in Appendix~\ref{sec:HSonfdC}). In the classical theory, physical processes are modeled by stochastic maps.
\item
Note that we are only demanding our \emph{prior} to be reproduced. If $\alpha'$ is some \emph{other} state on $\Alg{A}$, this axiom is \emph{not} saying that $\retro_{\alpha,\mathcal{E}}(\mathcal{E}(\alpha'))=\alpha'$. Note also that this axiom together with the first one allows us to think of retrodiction as an assignment that sends a pair $(\alpha,\mathcal{E})$ consisting of a valid state and process to $(\mathcal{E}(\alpha),\retro_{\alpha,\mathcal{E}})$, which is another valid state and process. This simplifies the form of the assignment since it can be viewed as a function that begins in one class of objects and comes back to that same class of objects. In this work, we will not consider time-reversals that do not satisfy this crucial state-preserving property, though we mention that some authors have considered dropping such an axiom~\cite{AZZ15}.
\item
Note that retrodicting on the identity process is, in particular, independent of the prior $\alpha$. 
\item
In this setting, we define $\mathcal{E}$ to be \emph{reversible} (i.e., \emph{invertible}) whenever there exists a process $\mathcal{E}^{-1}:\Alg{B}\to\Alg{A}$ such that $\mathcal{E}\circ\mathcal{E}^{-1}=\id_{\Alg{B}}$ and $\mathcal{E}^{-1}\circ\mathcal{E}=\id_{\Alg{A}}$. In the context of category theory, $\mathcal{E}$ is also called an \emph{isomorphism}.
\item
This axiom has appeared in many approaches towards time-reversal~\cite{AZZ15,Cr08,SSGC22}. A weakened form of involutivity was considered in~\cite{ChLi22}.
\item
This axiom and the next one describe in what sense retrodiction of a complicated process can be broken into retrodictions of simpler components. These are some of the essential axioms where the language of category theory becomes especially important. The compositional property was considered crucial in~\cite{ChLi22} and was also emphasized in many other works such as~\cite{Cr08}.
\item
In the tensorial axiom, we are assuming that our collection of states and processes has a tensorial structure, sometimes called \emph{parallel} composition to not conflate it with the \emph{series} composition from the previous axiom. This axiom seems to have been emphasized a bit less in the literature on time-reversal symmetry, but was emphasized in the context of recovery maps~\cite{LiWi18,JRSWW18,Wilde15}.
\end{enumerate}

\begin{figure}
\centering
\scalebox{0.5}{
\begin{tikzpicture}[scale=0.30]
  \draw[white,line width=0mm] (-35,-16) rectangle (35,19);
\def\nbx{4};
\def\nby{3};
\def\nbxs{9};
\node[blue] at (\nbx+6.5,\nby+5.5) {{\LARGE \textbf{bistochastic channels}}};
\draw[->,blue,ultra thick] (\nbx+4.5,\nby+2) -- (\nbx+8.5,\nby+2);
  \draw[blue,ultra thick,rounded corners=13mm] (-4,-13) rectangle (25,11);
      \coordinate (TRB) at (23,11);
     \draw[->,line width=1.2mm,color=blue] ($(TRB.south east)+(.4,0)$) node[above,xshift=-0.1cm,yshift=0.1cm]{{\Huge \textbf{adjoint}}} arc 
          [
               start angle=150,
               end angle=-85,
               radius=2.0cm
          ];
  	\draw[blue] (\nbx,\nby) rectangle (\nbx+4,\nby+4);
  	\fill[blue!50!white] (\nbx,\nby+3) rectangle (\nbx+1,\nby+4);
  	\fill[blue!50!white] (\nbx+1,\nby+2) rectangle (\nbx+2,\nby+3);
  	\fill[blue!50!white] (\nbx+2,\nby+1) rectangle (\nbx+3,\nby+2);
  	\fill[blue!50!white] (\nbx+3,\nby) rectangle (\nbx+4,\nby+1);
  	\draw[blue] (\nbx+\nbxs,\nby) rectangle (\nbx+4+\nbxs,\nby+4);
  	\fill[blue!50!white] (\nbx+\nbxs,\nby+3) rectangle (\nbx+1+\nbxs,\nby+4);
  	\fill[blue!50!white] (\nbx+1+\nbxs,\nby+2) rectangle (\nbx+2+\nbxs,\nby+3);
  	\fill[blue!50!white] (\nbx+2+\nbxs,\nby+1) rectangle (\nbx+3+\nbxs,\nby+2);
  	\fill[blue!50!white] (\nbx+3+\nbxs,\nby) rectangle (\nbx+4+\nbxs,\nby+1);
\def\nrx{4.5};
\def\nry{-4.5};
\def\nrxs{9};
\def\nrys{-0.5};
\node[red] at (\nrx+6.75,\nry-1.5) {{\LARGE \textbf{reversible channels}}};
\draw[->,red,ultra thick] (\nrx+4.5,\nry+2) -- (\nrx+8.5,\nry+2);
  \draw[red,ultra thick,rounded corners=10mm] (-2,-8) rectangle (22,1);
      \coordinate (BRR) at (19.5,-8.5);
     \draw[->,line width=1.2mm,color=red] ($(BRR.south east)+(.4,0)$) node[below,xshift=-0.7cm,yshift=-0.2cm]{{\Huge \textbf{adjoint}}} arc 
          [
               start angle=-150,
               end angle=65,
               radius=2.0cm
          ];
  	\draw[red] (\nrx,\nry) rectangle (\nrx+4,\nry+4);
  	\fill[red!30!white] (\nrx,\nry+3) rectangle (\nrx+1,\nry+4);
  	\fill[red!10!white] (\nrx+1,\nry+3) rectangle (\nrx+2,\nry+4);
  	\fill[red!50!white] (\nrx+2,\nry+3) rectangle (\nrx+3,\nry+4);
  	\fill[red!40!white] (\nrx+3,\nry+3) rectangle (\nrx+4,\nry+4);
  	\fill[red!10!white] (\nrx,\nry+2) rectangle (\nrx+1,\nry+3);
  	\fill[red!50!white] (\nrx+1,\nry+2) rectangle (\nrx+2,\nry+3);
  	\fill[red!60!white] (\nrx+2,\nry+2) rectangle (\nrx+3,\nry+3);
  	\fill[red!0!white] (\nrx+3,\nry+2) rectangle (\nrx+4,\nry+3);
  	\fill[red!50!white] (\nrx,\nry+1) rectangle (\nrx+1,\nry+2);
  	\fill[red!60!white] (\nrx+1,\nry+1) rectangle (\nrx+2,\nry+2);
  	\fill[red!30!white] (\nrx+2,\nry+1) rectangle (\nrx+3,\nry+2);
  	\fill[red!50!white] (\nrx+3,\nry+1) rectangle (\nrx+4,\nry+2);
  	\fill[red!40!white] (\nrx,\nry) rectangle (\nrx+1,\nry+1);
  	\fill[red!0!white] (\nrx+1,\nry) rectangle (\nrx+2,\nry+1);
  	\fill[red!50!white] (\nrx+2,\nry) rectangle (\nrx+3,\nry+1);
  	\fill[red!90!white] (\nrx+3,\nry) rectangle (\nrx+4,\nry+1);
  	\draw[red] (\nrx+\nrxs,\nry) rectangle (\nrx+4+\nrxs,\nry+4);
  	\fill[red!60!white] (\nrx+\nrxs,\nry+3) rectangle (\nrx+1+\nrxs,\nry+4);
  	\fill[red!30!white] (\nrx+1+\nrxs,\nry+3) rectangle (\nrx+2+\nrxs,\nry+4);
  	\fill[red!10!white] (\nrx+2+\nrxs,\nry+3) rectangle (\nrx+3+\nrxs,\nry+4);
  	\fill[red!20!white] (\nrx+3+\nrxs,\nry+3) rectangle (\nrx+4+\nrxs,\nry+4);
  	\fill[red!30!white] (\nrx+\nrxs,\nry+2) rectangle (\nrx+1+\nrxs,\nry+3);
  	\fill[red!70!white] (\nrx+1+\nrxs,\nry+2) rectangle (\nrx+2+\nrxs,\nry+3);
  	\fill[red!0!white] (\nrx+2+\nrxs,\nry+2) rectangle (\nrx+3+\nrxs,\nry+3);
  	\fill[red!60!white] (\nrx+3+\nrxs,\nry+2) rectangle (\nrx+4+\nrxs,\nry+3);
  	\fill[red!10!white] (\nrx+\nrxs,\nry+1) rectangle (\nrx+1+\nrxs,\nry+2);
  	\fill[red!0!white] (\nrx+1+\nrxs,\nry+1) rectangle (\nrx+2+\nrxs,\nry+2);
  	\fill[red!50!white] (\nrx+2+\nrxs,\nry+1) rectangle (\nrx+3+\nrxs,\nry+2);
  	\fill[red!30!white] (\nrx+3+\nrxs,\nry+1) rectangle (\nrx+4+\nrxs,\nry+2);
  	\fill[red!20!white] (\nrx+\nrxs,\nry) rectangle (\nrx+1+\nrxs,\nry+1);
  	\fill[red!60!white] (\nrx+1+\nrxs,\nry) rectangle (\nrx+2+\nrxs,\nry+1);
  	\fill[red!30!white] (\nrx+2+\nrxs,\nry) rectangle (\nrx+3+\nrxs,\nry+1);
  	\fill[red!20!white] (\nrx+3+\nrxs,\nry) rectangle (\nrx+4+\nrxs,\nry+1);
\def\nkx{-22};
\def\nky{6};
\def\nkxs{10};
\def\nkys{-0.5};
\node at (\nkx+7,\nky+5.5) {{\LARGE \textbf{quantum channels}}};
\draw[->,ultra thick] (\nkx+5.5,\nky+1.5) -- (\nkx+9.5,\nky+1.5);
  \draw[ultra thick,rounded corners=13mm] (-28,-15) rectangle (30,15);
    \coordinate (TLB) at (-26.5,15.25);
     \draw[->,line width=1.2mm] ($(TLB.north west)+(.4,0)$) node[left,xshift=-0.5cm,yshift=0.85cm]{{\Huge \textbf{Petz}}} arc 
          [
               start angle=25,
               end angle=260,
               radius=2.0cm
          ];
  	\draw[black] (\nkx,\nky-1) rectangle (\nkx+5,\nky+4);
  	\fill[black!30!white] (\nkx,\nky+3) rectangle (\nkx+1,\nky+4);
  	\fill[black!10!white] (\nkx+1,\nky+3) rectangle (\nkx+2,\nky+4);
  	\fill[black!50!white] (\nkx+2,\nky+3) rectangle (\nkx+3,\nky+4);
  	\fill[black!40!white] (\nkx+3,\nky+3) rectangle (\nkx+4,\nky+4);
  	\fill[black!20!white] (\nkx+4,\nky+3) rectangle (\nkx+5,\nky+4);
  	\fill[black!10!white] (\nkx,\nky+2) rectangle (\nkx+1,\nky+3);
  	\fill[black!50!white] (\nkx+1,\nky+2) rectangle (\nkx+2,\nky+3);
  	\fill[black!60!white] (\nkx+2,\nky+2) rectangle (\nkx+3,\nky+3);
  	\fill[black!0!white] (\nkx+3,\nky+2) rectangle (\nkx+4,\nky+3);
  	\fill[black!80!white] (\nkx+4,\nky+2) rectangle (\nkx+5,\nky+3);
  	\fill[black!50!white] (\nkx,\nky+1) rectangle (\nkx+1,\nky+2);
  	\fill[black!60!white] (\nkx+1,\nky+1) rectangle (\nkx+2,\nky+2);
  	\fill[black!30!white] (\nkx+2,\nky+1) rectangle (\nkx+3,\nky+2);
  	\fill[black!50!white] (\nkx+3,\nky+1) rectangle (\nkx+4,\nky+2);
  	\fill[black!10!white] (\nkx+4,\nky+1) rectangle (\nkx+5,\nky+2);
  	\fill[black!40!white] (\nkx,\nky) rectangle (\nkx+1,\nky+1);
  	\fill[black!0!white] (\nkx+1,\nky) rectangle (\nkx+2,\nky+1);
  	\fill[black!50!white] (\nkx+2,\nky) rectangle (\nkx+3,\nky+1);
  	\fill[black!70!white] (\nkx+3,\nky) rectangle (\nkx+4,\nky+1);
  	\fill[black!30!white] (\nkx+4,\nky) rectangle (\nkx+5,\nky+1);
  	\fill[black!20!white] (\nkx,\nky-1) rectangle (\nkx+1,\nky);
  	\fill[black!80!white] (\nkx+1,\nky-1) rectangle (\nkx+2,\nky);
  	\fill[black!10!white] (\nkx+2,\nky-1) rectangle (\nkx+3,\nky);
  	\fill[black!30!white] (\nkx+3,\nky-1) rectangle (\nkx+4,\nky);
  	\fill[black!20!white] (\nkx+4,\nky-1) rectangle (\nkx+5,\nky);
  	\draw[black] (\nkx+\nkxs,\nky+\nkys) rectangle (\nkx+4+\nkxs,\nky+4+\nkys);
  	\fill[black!60!white] (\nkx+\nkxs,\nky+3+\nkys) rectangle (\nkx+1+\nkxs,\nky+4+\nkys);
  	\fill[black!30!white] (\nkx+1+\nkxs,\nky+3+\nkys) rectangle (\nkx+2+\nkxs,\nky+4+\nkys);
  	\fill[black!10!white] (\nkx+2+\nkxs,\nky+3+\nkys) rectangle (\nkx+3+\nkxs,\nky+4+\nkys);
  	\fill[black!20!white] (\nkx+3+\nkxs,\nky+3+\nkys) rectangle (\nkx+4+\nkxs,\nky+4+\nkys);
  	\fill[black!30!white] (\nkx+\nkxs,\nky+2+\nkys) rectangle (\nkx+1+\nkxs,\nky+3+\nkys);
  	\fill[black!70!white] (\nkx+1+\nkxs,\nky+2+\nkys) rectangle (\nkx+2+\nkxs,\nky+3+\nkys);
  	\fill[black!0!white] (\nkx+2+\nkxs,\nky+2+\nkys) rectangle (\nkx+3+\nkxs,\nky+3+\nkys);
  	\fill[black!60!white] (\nkx+3+\nkxs,\nky+2+\nkys) rectangle (\nkx+4+\nkxs,\nky+3+\nkys);
  	\fill[black!10!white] (\nkx+\nkxs,\nky+1+\nkys) rectangle (\nkx+1+\nkxs,\nky+2+\nkys);
  	\fill[black!0!white] (\nkx+1+\nkxs,\nky+1+\nkys) rectangle (\nkx+2+\nkxs,\nky+2+\nkys);
  	\fill[black!50!white] (\nkx+2+\nkxs,\nky+1+\nkys) rectangle (\nkx+3+\nkxs,\nky+2+\nkys);
  	\fill[black!30!white] (\nkx+3+\nkxs,\nky+1+\nkys) rectangle (\nkx+4+\nkxs,\nky+2+\nkys);
  	\fill[black!20!white] (\nkx+\nkxs,\nky+\nkys) rectangle (\nkx+1+\nkxs,\nky+1+\nkys);
  	\fill[black!60!white] (\nkx+1+\nkxs,\nky+\nkys) rectangle (\nkx+2+\nkxs,\nky+1+\nkys);
  	\fill[black!30!white] (\nkx+2+\nkxs,\nky+\nkys) rectangle (\nkx+3+\nkxs,\nky+1+\nkys);
  	\fill[black!20!white] (\nkx+3+\nkxs,\nky+\nkys) rectangle (\nkx+4+\nkxs,\nky+1+\nkys);
\def\nox{-21};
\def\noy{-5};
\def\noxs{10};
\def\noys{-0.5};
\node[violet] at (\nox+7,\noy-1.5) {{\LARGE \textbf{classical channels}}};
\draw[->,violet,ultra thick] (\nox+5.5,\noy+2.5) -- (\nox+9.5,\noy+2.5);
  \draw[violet,ultra thick,rounded corners=13mm] (-24,-10) rectangle (1,3); 
      \coordinate (BLV) at (-22.5,-10.5);
     \draw[->,line width=1.2mm,color=violet] ($(BLV.south west)+(.4,0)$) node[below,xshift=0.3cm,yshift=-0.3cm]{{\Huge \textbf{Bayes}}} arc 
          [
               start angle=-25,
               end angle=-260,
               radius=2.0cm
          ];
  	\draw[violet] (\nox,\noy) rectangle (\nox+5,\noy+5);
  	\fill[violet!40!white] (\nox,\noy+4) rectangle (\nox+1,\noy+5);
  	\fill[violet!20!white] (\nox+1,\noy+3) rectangle (\nox+2,\noy+4);
  	\fill[violet!60!white] (\nox+2,\noy+2) rectangle (\nox+3,\noy+3);
  	\fill[violet!30!white] (\nox+3,\noy+1) rectangle (\nox+4,\noy+2);
  	\fill[violet!50!white] (\nox+4,\noy) rectangle (\nox+5,\noy+1);
  	\draw[violet] (\nox+\noxs,\noy+\noys+1) rectangle (\nox+5+\noxs-1,\noy+5+\noys);
  	\fill[violet!60!white] (\nox+\noxs,\noy+4+\noys) rectangle (\nox+1+\noxs,\noy+5+\noys);
  	\fill[violet!40!white] (\nox+1+\noxs,\noy+3+\noys) rectangle (\nox+2+\noxs,\noy+4+\noys);
  	\fill[violet!30!white] (\nox+2+\noxs,\noy+2+\noys) rectangle (\nox+3+\noxs,\noy+3+\noys);
  	\fill[violet!70!white] (\nox+3+\noxs,\noy+1+\noys) rectangle (\nox+4+\noxs,\noy+2+\noys);
\end{tikzpicture}
}
\caption{The 
sets (not drawn to scale) and their inclusion structure depict four families of channels (the inclusion structure is not meant to include the states). The standard time-reversal symmetry is obtained by taking the Hilbert--Schmidt adjoint of a reversible channel, or, more generally, a bistochastic channel. The states (drawn schematically as matrices with shaded entries) are irrelevant for reversible channels, but are implicitly 
the uniform states for bistochastic channels. For classical channels equipped with arbitrary probability distributions, standard Bayesian inversion provides a form of time-reversal symmetry that goes beyond the Hilbert--Schmidt adjoint for bistochastic channels. Finally, the Petz recovery map allows an extension of Bayesian inversion to all quantum channels and arbitrary states. In brief, this paper isolates axioms for retrodiction and inferential time-reversal symmetry that are simultaneously satisfied by all of these classes of channels and states.}
\label{fig:trpetz}
\end{figure}
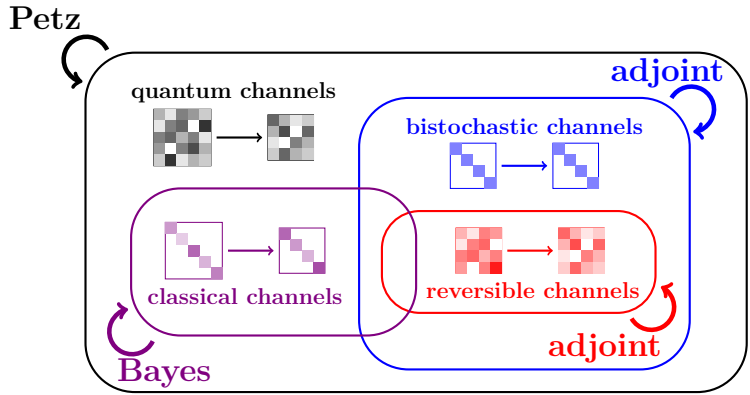

In this paper (specifically Theorem~\ref{thm:involutivePetz}), we prove that among a variety of different proposals for recovery maps and retrodiction in the quantum setting (including Petz recovery maps, their rotated variants, and their averaged generalizations, and many other candidates), the only one that satisfies all these axioms is the Petz recovery map~\cite{Pe88,BaKn02,Le06,Le07,Cr08}. This justifies the Petz recovery map as a retrodiction map, and hence an extension of time-reversal symmetry to all quantum channels (see Figure~\ref{fig:trpetz}). 
Although we do not \emph{characterize} the Petz recovery map (and hence Bayesian inversion in the classical setting) among \emph{all} possible retrodiction maps, we propose a precise mathematical problem whether our axioms indeed isolate Petz among \emph{all} possible retrodiction maps (and not just the vast examples we have listed).

We emphasize that our axioms are logical as opposed to analytical. It has often been the case that axioms used to help single out a form of retrodiction (a recovery map to be more precise) optimized some quantity, such as the difference of relative entropies, more general divergences, fidelity of recovery, or relative entropy of recovery~\cite{Pe88,BaKn02,Pe03,NgMa10,FaRe15,Je17,LiWi18,SuSc22}. Rather than choosing a specific such distance measure and then finding axioms to argue for the necessity of those, we have preferred to find a logical set of axioms more directly, in spirit of earlier derivations of classical inference~\cite{Cs91}. In addition, the axioms that we identified do not involve concepts such as measurement, observations, or the need of any intervention, which are concepts that are mathematically difficult to define outside of specific models. In fact, our axioms can be formulated without any mention of probability theory. Our axioms are merely those of consistency rather than properties we might expect from our experiences, which are largely based on classical thinking, and may therefore skew our understanding of quantum. 

Furthermore, we illustrate in detail which axioms fail for various other proposals of retrodiction by providing explicit examples. In particular, we illustrate, to some degree, how independent most of our axioms are. For example, the averaged rotated Petz recovery maps that have appeared recently in the context of strengthening data-processing inequalities via recovery maps~\cite{JRSWW16,JRSWW18} are \emph{neither} compositional, tensorial, nor involutive. Furthermore, we show that the recent proposal of Surace--Scandi~\cite{SuSc22} on state-retrieval maps is also not compositional. This emphasizes some of the key differences between retrodiction and approximate error correction~\cite{BaKn02,NgMa10,SuSc22}, and this distinction may have important consequences for quantum information in extreme situations, such as near the horizons of black holes, where state-dependent approximate error-correction has recently been used to suggest that information might be stored in Hawking radiation in certain models of black hole evaporation~\cite{AEMM19,CHPSSW19,Pe20,CPS20,No21,AkPe22,AEHPV22}.

\section{Retrodiction as a monoidal functor}

In the setting of quantum theory, we would like to define retrodiction as an assignment that takes a prior, defined on some matrix algebra, together with a process involving another matrix algebra, and produces a map in the opposite direction. However, working with matrix algebras is unnecessarily restrictive and it is more appropriate to include classical and hybrid classical/quantum systems in our analysis. 
As such, we will model our \emph{systems} with finite-dimensional unital $C^*$-algebras, where commutative algebras correspond to classical systems and general non-commutative algebras correspond to quantum (or hybrid) systems. 
We will model our \emph{processes} with completely positive trace-preserving (CPTP) maps between such $C^*$-algebras. 
Briefly, one benefit of using $C^*$-algebras, as opposed to only matrix algebras, is that one can express quantum channels, classical stochastic maps, measurements, preparations, and instruments all as processes between $C^*$-algebras (see~\cite[Section~II.A]{FuPa22QB} for more details). 
We are expressing processes in the Schr\"odinger picture where the processes describe the action on states. 
We emphasize that we do \emph{not} require $\mathcal{E}$ to preserve any of the algebraic structure, nor do we require the unit of the algebras to be preserved.
The reader should not be discouraged by our usage of finite-dimensional $C^*$-algebras because they are equivalent to direct sums of matrix algebras~\cite[Theorem~5.20 and Proposition~5.26]{Fa01}. We henceforth take the convention that all $C^*$-algebras appearing in this work will be finite dimensional and unital unless specified otherwise. 

We now introduce a category that mathematically describes priors and processes.

\bd
The \define{category of faithful states} is the category $\St$ whose objects consists of pairs $(\Alg{A},\alpha)$, where $\Alg{A}$ is a finite-dimensional unital $C^*$-algebra and $\alpha$ is a \define{faithful state} on $\Alg{A}$, i.e., $\alpha$ is positive, $\tr(\alpha)=1$, and $\alpha$ is non-degenerate in the sense that if $A\in\Alg{A}$ satisfies $\tr(\alpha A^{\dag}A)=0$, then $A=0$. A morphism from $(\Alg{A},\alpha)$ to $(\Alg{B},\beta)$ in $\St$ is a CPTP map
 $\Alg{A}\xrightarrow{\mathcal{E}}\Alg{B}$ such that $\mathcal{E}(\alpha)=\beta$. Such a morphism is drawn as $(\Alg{A},\alpha)\xrightarrow{\mathcal{E}}(\Alg{B},\beta)$. The composition in $\St$ is the composition of functions and the identity morphism on $(\Alg{A},\alpha)$ is the identity $\id_{\Alg{A}}$ map on $\Alg{A}$.%
\footnote{The category $\St$ is not quite the coslice/under category $\C\downarrow\CPTP$, where $\CPTP$ is the category of finite-dimensional $C^*$-algebras and CPTP maps, but rather a subcategory where all states are faithful.}
Let $\CSt$ denote the subcategory of faithful states on \emph{commutative} $C^*$-algebras. 
\ed

We leave the verification that the axioms of a category hold for $\St$ and $\CSt$ to the reader.
As described in the introduction, $(\Alg{A},\alpha)$ represents a physical system, represented by an algebra $\Alg{A}$, together with a faithful state $\alpha$ on $\Alg{A}$. A morphism $(\Alg{A},\alpha)\xrightarrow{\mathcal{E}}(\Alg{B},\beta)$ acquires the physical interpretation that $\alpha$ is a \emph{prior} and the process $\mathcal{E}$ takes this prior to $\beta$, written as $\beta=\mathcal{E}(\alpha)$. As such, $\beta$ is called the \emph{prediction}. 

The usage of faithful states as opposed to all states has at least two purposes. One reason is for mathematical simplicity, since subtle issues arise when dealing with non-faithful states, which allows for states to produce expectation values of $0$ on certain positive operators. We believe that focusing on faithful states first prevents us from being distracted by the technical issues associated with measure zero, where we need to discuss projections and their orthogonal complements, substantially lengthening many of the formulas and calculations.  The second reason is to allow ourselves to not completely rule out events that we believe are impossible. It is conceivable that some events may happen with a small enough probability that it is \emph{effectively} vanishing, but technically \emph{non-vanishing}. For these two reasons, we will focus exclusively on faithful states in this work. As such, all states will be faithful from now on unless specified otherwise. 

Having introduced the main category of discussion, we now come to explicating some of the basic axioms that we believe retrodiction should satisfy. First, if one is given a morphism $(\Alg{A},\alpha)\xrightarrow{\mathcal{E}}(\Alg{B},\beta)$ describing a one step dynamics evolution involving a prior $\alpha$ with $\beta$ as the prediction, then the retrodiction of this morphism should be a CPTP map of the form 
\[
\Alg{A}\xleftarrow{\retro_{\alpha,\mathcal{E}}}\Alg{B}
\]
reversing the directionality of the original process $\Alg{A}\xrightarrow{\mathcal{E}}\Alg{B}$. Note that we are only including $\alpha$ and $\mathcal{E}$ in the notation for the retrodiction since the prediction $\beta=\mathcal{E}(\alpha)$ can be obtained from these data. 
Also note that we have only motivated that $\retro_{\alpha,\mathcal{E}}$ should be positive and trace-preserving, rather than CPTP, because we want to ensure that states are always mapped to states. It is only after we introduce the tensor product structure later that we will see the usefulness of assuming retrodiction to be CPTP.

Furthermore, the retrodiction map $\retro_{\alpha,\mathcal{E}}$ should \emph{preserve states} in the sense that it sends the prediction $\beta$ back to the prior $\alpha$, i.e., $\retro_{\alpha,\mathcal{E}}(\beta)=\alpha$, or more suggestively $\retro_{\alpha,\mathcal{E}}(\mathcal{E}(\alpha))=\alpha$. Note, however, that this does \emph{not} imply that $\retro_{\alpha,\mathcal{E}}(\mathcal{E}(\alpha'))=\alpha'$ for any \emph{other} $\alpha'$ different from $\alpha$. Such a condition would drastically reduce the collection of maps $\mathcal{E}$ that would be of interest to us because they would necessarily be invertible. This is discussed later in Theorem~\ref{thm:AwBuSc}. One can interpret $\retro_{\alpha,\mathcal{E}}$ acting on an \emph{arbitrary} state $\beta\in\Alg{B}$ as a quantum generalization of Jeffrey's \emph{probability kinematics}~\cite{Je90} (see Proposition~\ref{prop:classicalBayes} and Proposition~\ref{prop:Petzimpliesbayes}).

Combining the requirements discussed in the previous paragraphs precisely say that 
\[
(\Alg{A},\alpha)\xleftarrow{\retro_{\alpha,\mathcal{E}}}(\Alg{B},\beta).
\]
should be a morphism in $\St$. As such, retrodiction should be some assignment of the form%
\footnote{If $\mathbf{C}$ is a category, $\mathbf{C}^{\op}$ denotes its opposite. By definition, a morphism from $\Alg{A}$ to $\Alg{B}$ in $\mathbf{C}^{\op}$ is a morphism from $\Alg{B}$ to $\Alg{A}$ in $\mathbf{C}$.}
\[
\St\xrightarrow{\retro}\St^{\op},
\]
where an object $(\Alg{A},\alpha)$ is sent to itself, while a morphism $(\Alg{A},\alpha)\xrightarrow{\mathcal{E}}(\Alg{B},\beta)$ is sent to a morphism $(\Alg{A},\alpha)\xleftarrow{\retro_{\alpha,\mathcal{E}}}(\Alg{B},\beta)$. This motivates the following definition (if $\mathbf{C}$ denotes a category, and if $a$ and $b$ are two objects in $\mathbf{C}$, then $\mathbf{C}(a,b)$ denotes the set of morphisms from $a$ to $b$). 

\bd
A \define{retrodiction family} is an assignment $\retro:\St\to\St^{\op}$
that acts as the identity on objects. Explicitly, this means that for every pair of objects $(\Alg{A},\alpha)$ and $(\Alg{B},\beta)$ in $\St$, $\retro$ defines a function $\St\big((\Alg{A},\alpha),(\Alg{B},\beta)\big)\to\St\big((\Alg{B},\beta),(\Alg{A},\alpha)\big)$ sending a morphism $(\Alg{A},\alpha)\xrightarrow{\mathcal{E}}(\Alg{B},\beta)$ to a morphism $(\Alg{A},\alpha)\xleftarrow{\retro_{\alpha,\mathcal{E}}}(\Alg{B},\beta)$.
\ed

We would like retrodiction to satisfy a few more intuitive axioms than just these basic ones. In particular, we have not required a retrodiction family to be a functor. As of now, the only axioms we have incorporated are that the prior is recovered and that retrodiction remains CPTP. We have also assumed the axiom of \emph{universality}, which says that retrodiction is a function depending only on the initial prior and the process (and nothing more). 

One other natural axiom to assume for retrodiction is that if one considers the \emph{identity} morphism $(\Alg{A},\alpha)\xrightarrow{\id_{\Alg{A}}}(\Alg{A},\alpha)$ in $\St$ for any state $\alpha$ on $\Alg{A}$, then the retrodiction of this should be the identity itself. As an equation, this can be expressed as $\retro_{\alpha,\id_{\Alg{A}}}=\id_{\Alg{A}}$. In words, retrodicting on the process that does nothing is itself the process that does nothing. This axiom is sometimes called \emph{normalization} in the literature~\cite{LiWi18,Wilde15,JRSWW18}. 

A second axiom that we believe retrodiction should satisfy is \emph{compositionality}. Namely, if one has two successive processes together with a prior (and successive predictions)
\[
(\Alg{A},\alpha)\xrightarrow{\mathcal{E}}(\Alg{B},\beta)\xrightarrow{\mathcal{F}}(\Alg{C},\gamma),
\]
then one can construct \emph{three} retrodictions associated with these data. On the one hand, one can retrodict on the individual processes $\mathcal{E}$ and $\mathcal{F}$ to construct two successive retrodictions 
\[
(\Alg{A},\alpha)\xleftarrow{\retro_{\alpha,\mathcal{E}}}(\Alg{B},\beta)\xleftarrow{\retro_{\beta,\mathcal{F}}}(\Alg{C},\gamma).
\]
On the other hand, one can construct the retrodiction associated with the \emph{composite} process
\[
(\Alg{A},\alpha)\xrightarrow{\mathcal{F}\circ\mathcal{E}}(\Alg{C},\gamma)
\]
in which case one obtains 
\[
(\Alg{A},\alpha)\xleftarrow{\retro_{\alpha,\mathcal{F}\circ\mathcal{E}}}(\Alg{C},\gamma).
\]
The axiom of \emph{compositionality} states that the composite of retrodictions for the two individual processes should equal the retrodiction of the composite process (cf.\ \cite{Cr08,LiWi18,Wilde15,JRSWW18}), i.e.,%
\footnote{If we excluded the prior from our notation for retrodiction, then this would look like $\retro_{\mathcal{F}\circ\mathcal{E}}=\retro_{\mathcal{E}}\circ\retro_{\mathcal{F}}$. We have chosen to include the prior in the notation to be more consistent with the quantum information literature.}
\[
\retro_{\alpha,\mathcal{F}\circ\mathcal{E}}=\retro_{\alpha,\mathcal{E}}\circ\retro_{\beta,\mathcal{F}}. 
\]

Thus, normalization and compositionality precisely say that retrodiction should be a functor, written $\St\xrightarrow{\retro}\St^{\op}$. 

But there is another structure we would like retrodiction to preserve, and this is the natural tensor product structure for states. Indeed, $\St$ and $\CSt$ are (symmetric) monoidal categories, where the tensor product of two objects $(\Alg{A},\alpha)$ and $(\Alg{A}',\alpha')$ is $(\Alg{A}\otimes\Alg{A}',\alpha\otimes\alpha')$. The tensor product of two morphisms $(\Alg{A},\alpha)\xrightarrow{\mathcal{E}}(\Alg{B},\beta)$ and $(\Alg{A}',\alpha')\xrightarrow{\mathcal{E}'}(\Alg{B}',\beta')$ is the usual tensor product of CPTP maps, namely 
\[
(\Alg{A}\otimes\Alg{A}',\alpha\otimes\alpha')\xrightarrow{\mathcal{E}\otimes\mathcal{E}'}(\Alg{B}\otimes\Alg{B}',\beta\otimes\beta').
\]
Intuitively, this monoidal (i.e., tensor product) structure on $\St$ (and $\CSt$) says that if two systems are prepared independently and undergo independent evolution, then the joint prior and joint process is also a valid prior and process. Due to this independence, we would therefore also expect that if we independently retrodict on each process via
\[
(\Alg{A},\alpha)\xleftarrow{\retro_{\alpha,\mathcal{E}}}(\Alg{B},\beta)
\qquad\text{ and }\qquad
(\Alg{A}',\alpha')\xleftarrow{\retro_{\alpha',\mathcal{E}'}}(\Alg{B}',\beta'),
\]
then the joint retrodiction
\[
(\Alg{A}\otimes\Alg{A}',\alpha\otimes\alpha')\xleftarrow{\retro_{\alpha,\mathcal{E}}\otimes\retro_{\alpha',\mathcal{E}'}}(\Alg{B}\otimes\Alg{B}',\beta\otimes\beta'),
\] 
should be the same as retrodicting on the joint process itself
\[
(\Alg{A}\otimes\Alg{A}',\alpha\otimes\alpha')\xleftarrow{\retro_{\alpha\otimes\alpha',\mathcal{E}\otimes\mathcal{E}'}}(\Alg{B}\otimes\Alg{B}',\beta\otimes\beta'),
\]
i.e., we expect retrodiction to satisfy \emph{tensoriality}, which states
\[
\retro_{\alpha,\mathcal{E}}\otimes\retro_{\alpha',\mathcal{E}'}=\retro_{\alpha\otimes\alpha',\mathcal{E}\otimes\mathcal{E}'}.
\]

The mathematical way to summarize the above discussion is to say that retrodiction should be, at the very least, a monoidal functor $\St\to\St^{\op}$
that acts as the identity on objects (we will also require other properties for retrodiction later).
Hence, already from these basic requirements of retrodiction, we are inevitably led to a description of retrodiction in terms of the mathematical discipline of category theory, specifically that of (symmetric) monoidal categories and monoidal functors. Note that these mathematical structures also appear in the settings of resource theories~\cite{CFS16}, topological field theories~\cite{At90,Fr13}, topological order~\cite{Ko14,BBCW19}, and many other contexts relevant to physics~\cite{BaSt11}.  

Before moving on to more reasonable examples of retrodiction, we first examine a non-example that comes quite close.

\bn
The assignment $\retro^{\shriek}:\St\to\St^{\op}$ given by sending a morphism $(\Alg{A},\alpha)\xrightarrow{\mathcal{E}}(\Alg{B},\beta)$ to the composite%
\footnote{The map $\C\xrightarrow{\td\alpha}\Alg{A}$ is the unique linear map determined by setting $\td\alpha(1)=\alpha$.}
\[
\begin{split}
(\Alg{A},\alpha)\xleftarrow{\td\alpha}(\C,1)&\xleftarrow{\tr}(\Alg{B},\beta)\\
\tr(B)\alpha\mapsfrom\tr(B)&\mapsfrom B
\end{split}
\]
is a retrodiction family that is compositional and tensorial. However, it does not satisfy the normalization property and is therefore not a functor.
It is called the \define{discard-and-prepare retrodiction family}.
\en

\bprf
First note that $\retro^{\shriek}_{\alpha,\mathcal{E}}=\td\alpha\circ\tr$ is CPTP because it is the composite of two CPTP maps. By construction, it preserves the states since $\retro^{\shriek}_{\alpha,\mathcal{E}}(\beta)=\td\alpha\big(\tr(\beta)\big)=\td\alpha(1)=\alpha$. Thus, $\retro^{\shriek}$ is a retrodiction family. 
To see that $\retro^{\shriek}$ is compositional, let $(\Alg{A},\alpha)\xrightarrow{\mathcal{E}}(\Alg{B},\beta)\xrightarrow{\mathcal{F}}(\Alg{C},\gamma)$ be a composable triple in $\St$. Then
\[
\retro^{\shriek}_{\alpha,\mathcal{E}}\circ\retro^{\shriek}_{\beta,\mathcal{F}}
=(\td\alpha\circ\tr)\circ(\td\beta\circ\tr)
=\td\alpha\circ\id_{\C}\circ\tr
=\td\alpha\circ\tr
=\retro^{\shriek}_{\alpha,\mathcal{F}\circ\mathcal{E}}
\]
proves compositionality. To see that $\retro^{\shriek}$ is tensorial, let $(\Alg{A},\alpha)\xrightarrow{\mathcal{E}}(\Alg{B},\beta)$ and $(\Alg{A}',\alpha')\xrightarrow{\mathcal{E}'}(\Alg{B}',\beta')$ be a pair of morphisms in $\St$. Then 
\[
\retro^{\shriek}_{\alpha,\mathcal{E}}\otimes\retro^{\shriek}_{\alpha',\mathcal{E}'}
=(\td\alpha\circ\tr)\otimes(\td\alpha'\circ\tr)
=(\td\alpha\otimes\td\alpha')\circ(\tr\otimes\tr)
=(\td\alpha\otimes\td\alpha')\circ\tr
=\retro^{\shriek}_{\alpha\otimes\alpha',\mathcal{E}\otimes\mathcal{E}'}
\]
proves tensoriality. Finally, $\retro^{\shriek}$ does not satisfy normalization since given any algebra $\Alg{A}$ whose dimension is greater than $1$, one obtains $\retro^{\shriek}_{\alpha,\id_{\Alg{A}}}=\td\alpha\circ\tr$ which is \emph{never} equal to $\id_{\Alg{A}}$ since the linear rank of this map is $1$. 
\eprf

It is worthwhile to briefly restrict attention to the classical case before moving on to more complicated retrodiction families in the fully quantum setting. First recall that any commutative $C^*$-algebra is $*$-isomorphic to one of the form $\C^{X}:=\bigoplus_{x\in X}\C$ for some finite set $X$. A state $p$ on $\C^{X}$ is given by a probability $\{p_{x}\}$ on $X$. A morphism $(\C^{X},p)\xrightarrow{\mathcal{E}}(\C^{Y},q)$ is uniquely determined by a stochastic matrix $\{\mathcal{E}_{yx}\}$ whose $yx$ entry describes the conditional probability of $y$ given $x$, i.e., $\mathcal{E}_{yx}\ge0$ for all $x\in X$, $y\in Y$, and $\sum_{y\in Y}\mathcal{E}_{yx}=1$ for all $x\in X$. 

\bn
\label{prop:classicalBayes}
The assignment
$\retro^{\Bayes}:\CSt\to\CSt^{\op}$ 
sending $(\C^{X},p)\xrightarrow{\mathcal{E}}(\C^{Y},q)$ to $(\C^{X},p)\xleftarrow{\overline{\mathcal{E}}:=\retro^{\Bayes}_{p,\mathcal{E}}}(\C^{Y},q)$, determined by the formula
\[
\overline{\mathcal{E}}_{xy}:=\frac{\mathcal{E}_{yx}p_{x}}{q_{y}}
\]
for all $x\in X$ and $y\in Y$, defines a monoidal functor that acts as the identity on objects. It is called \define{Bayesian inversion} or the \define{Bayesian retrodiction family}. 
\en

Note that $\retro^{\Bayes}_{\alpha,\mathcal{E}}:\C^{Y}\to\C^{X}$ acting on the vector $\delta_{y}\in\C^{Y}$, whose value is $1$ at $y$ and $0$ elsewhere, gives the probability vector $\retro^{\Bayes}_{\alpha,\mathcal{E}}(\delta_{y})$ whose $x^{\text{th}}$ component is $\overline{\mathcal{E}}_{xy}$. This reproduces Bayes' update rule based on obtaining hard evidence $y$. More generally, $\retro^{\Bayes}_{\alpha,\mathcal{E}}$ acting on an \emph{arbitrary} probability $r\in\C^{Y}$ gives the probability vector $\retro^{\Bayes}_{\alpha,\mathcal{E}}(r)$ whose $x^{\text{th}}$ component is $\sum_{y\in Y}\overline{\mathcal{E}}_{xy}r_{y}$. This reproduces Jeffrey's update rule associated with soft evidence represented by the probability $r$ (also called \emph{probability kinematics})~\cite{Pearl88,Je90,Ja03,Ja19}. 
A large part of this paper is devoted to finding an extension of $\retro^{\Bayes}$ from $\CSt$ to all of $\St$ to include quantum channels as well. In other words, our goal is to find a retrodiction family that extends classical retrodiction defined in terms of Bayes' and Jeffrey's update rule. 

Importantly, note that we do not impose the condition that a retrodiction family $\retro:\St\to\St^{\op}$ agrees with classical Bayesian inversion when restricted to $\CSt$. We suspect, though we do not have a proof at present, that classical Bayesian inversion is characterized by such functoriality properties (see Question~\ref{que:retrocharacter} for more details).

\section{The Petz recovery maps}

One immediately wonders if any retrodiction families that are monoidal functors actually exist, how many there are, and if we can find an explicit form for all of them. The first question was answered in~\cite{LiWi18}, and a detailed proof was provided in~\cite{Wilde15} (though the phrasing and emphasis were different), though earlier partial results were also obtained in~\cite{Cr08}. However, we will prove more general results in this work, so we will restate these results using our language. In what follows, we will show that any rotated Petz recovery map can be used to define a retrodiction family that is a monoidal functor. However, \emph{averaged} rotated Petz recovery maps in general do not satisfy compositionality and tensoriality, and this will be expounded on in later sections. To establish these claims, we first recall some definitions. 

\bd
\label{defn:Petzmaps}
Let $(\Alg{A},\alpha)\xrightarrow{\mathcal{E}}(\Alg{B},\beta)$ be a morphism in $\St$. The \define{Petz recovery map} associated with the pair $(\alpha,\mathcal{E})$ is the morphism%
\footnote{The reader is expected to check that this morphism indeed preserves the states and is a CPTP map. The reader is invited to look at~\cite{Wilde15} for a simple proof.}
\[
(\Alg{A},\alpha)\xleftarrow{\retro^{\Petz}_{\alpha,\mathcal{E}}}(\Alg{B},\beta)
\]
in $\St$ defined by the formula
\[
\retro^{\Petz}_{\alpha,\mE}
:=\Ad_{\alpha^{1/2}}\circ \mE^*\circ\Ad_{\beta^{-1/2}}
,
\]
where $\mE^*$ is the Hilbert--Schmidt adjoint of $\mE$ (cf.\ Appendix~\ref{sec:HSonfdC}) and $\Ad_{V}$ is the map sending $A$ to $VAV^{\dag}$. 
For any $t\in\R$, the \define{rotated Petz recovery map} associated with  $\mE$ is the morphism%
\footnote{Again, the reader is expected to check that this is indeed a morphism in $\St$.}
\[
(\Alg{A},\alpha)\xleftarrow{\retro^{\Petz,t}_{\alpha,\mE}}(\Alg{B},\beta)
\]
in $\St$ defined by the formula 
\[
\retro^{\Petz,t}_{\alpha,\mE}:=\Ad_{\alpha^{-it}}\circ\retro^{\Petz}_{\alpha,\mE}\circ\Ad_{\beta^{it}}
\equiv
\Ad_{\alpha^{1/2-it}}\circ\mE^*\circ\Ad_{\beta^{-1/2+it}}.
\]
Finally, for any probability measure $\mu$ on $\R$, the \define{$\mu$-averaged rotated Petz recovery map} associated with the pair $(\alpha,\mE)$ is the morphism%
\footnote{We will prove that this is indeed a morphism in $\St$ in Proposition~\ref{prop:RPmupreretro}.}
\[
(\Alg{A},\alpha)\xleftarrow{\retro^{\Petz,\mu}_{\alpha,\mE}}(\Alg{B},\beta)
\]
in $\St$ defined by the formula 
\[
\retro^{\Petz,\mu}_{\alpha,\mE}(B):=\int_{-\infty}^{\infty}\retro^{\Petz,t}_{\alpha,\mE}(B)\,d\mu(t)
\equiv
\int_{-\infty}^{\infty}\left(\Ad_{\alpha^{1/2-it}}\circ\mE^*\circ\Ad_{\beta^{-1/2+it}}\right)(B)\,d\mu(t).
\]
\ed

An example of an averaged rotated Petz recovery map is one given by a finite convex combination of rotated Petz recovery maps at different values of $t\in\R$ (see Appendix~\ref{sec:bitflipavgPetz} for an explicit example). Yet another example is one that has recently appeared in the literature as a universal recovery map when strengthening the data-processing inequality~\cite{JRSWW16,JRSWW18}. 

\bx
\label{ex:JRSWW}
The \define{Junge--Renner--Sutter--Wilde--Winter (JRSWW) recovery map} $\retro^{\JRSWW}$ is the averaged rotated Petz recovery map $\retro^{\Petz,\mu}$ with probability measure $\mu$ defined by the density%
\footnote{The slight difference between our formula and the one in~\cite{JRSWW16,JRSWW18} is because of our insistence of using $\retro^{\Petz,t}$ inside the integral, whereas~\cite{JRSWW16,JRSWW18} use $\retro^{\Petz,\frac{t}{2}}$.}
\[
d\mu(t)=\pi\big(\cosh(2\pi t)+1\big)^{-1} dt.
\]
\ex

We now come to describing the above examples in terms of retrodiction families.

\bt
\label{thm:RPtretro}
For every $t\in\R$, the assignment 
$\retro^{\Petz,t}:\St\to\St^{\op}$ 
sending a morphism $(\Alg{A},\alpha)\xrightarrow{\mE}(\Alg{B},\beta)$ to $(\Alg{A},\alpha)\xleftarrow{\retro^{\Petz,t}_{\alpha,\mE}}(\Alg{B},\beta)$
defines a retrodiction family that is a monoidal functor. 
\et

\bprf
Modulo the fact that we are working with arbitrary finite-dimensional $C^*$-algebras, as opposed to just matrix algebras, this is proved in~\cite{Wilde15}. More specifically, \cite[Remark~3.5]{Wilde15} proves the state-preservation condition, \cite[Appendix~B]{Wilde15} proves $\retro^{\Petz,t}_{\alpha,\mE}$ is CPTP (since the projection is the identity by our assumption of faithfulness for states), while the normalization, compositionality, and tensoriality properties are proved in~\cite[Section~4]{Wilde15}. The same proof works for arbitrary finite-dimensional $C^*$-algebras.
\eprf

\bn
\label{prop:RPmupreretro}
Let $\mu$ be a probability measure on $\R$. Then 
$\retro^{\Petz,\mu}:\St\to\St^{\op}$
sending a morphism $(\Alg{A},\alpha)\xrightarrow{\mE}(\Alg{B},\beta)$ to $(\Alg{A},\alpha)\xleftarrow{\retro^{\Petz,\mu}_{\alpha,\mE}}(\Alg{B},\beta)$
defines a retrodiction family that satisfies normalization. 
\en

\bprf
The state-preservation condition follows from 
\[
\retro^{\Petz,\mu}_{\alpha,\mE}(\beta)
=\int_{-\infty}^{\infty}\retro^{\Petz,t}_{\alpha,\mE}(\beta)\;d\mu(t)
=\int_{-\infty}^{\infty}\alpha\;d\mu(t)
=\alpha\int_{-\infty}^{\infty}d\mu(t)
=\alpha, 
\]
where we have used the state-preservation property of the rotated Petz recovery map in the second equality. 
The fact that $\retro^{\Petz,\mu}$ is CP follows from the fact that it can be expressed as an increasing limit of positive linear combinations of CP maps. The fact that $\retro^{\Petz,\mu}$ is trace-preserving is a consequence of the fact that $\retro^{\Petz,t}$ is trace-preserving for all $t\in\R$ and since $\mu$ is a probability measure. 
Finally, $\retro^{\Petz,\mu}$ satisfies the normalization condition for the same reason, namely because $\retro^{\Petz,t}_{\alpha,\id_{\Alg{A}}}=\id_{\Alg{A}}$ for all $t\in\R$ and because $\mu$ is a probability measure. 
\eprf

There is a generalization of rotated Petz recovery maps where \emph{independent} phases act on the eigenspaces of the corresponding states. These were used to provide a strengthening of the data-processing inequality by Sutter, Tomamichel, and Harrow (STH)~\cite{SuToHa16}. 

\bd
\label{defn:STH}
Let $\alpha$ and $\beta$ be (faithful) states on 
$C^*$-algebras $\Alg{A}$ and $\Alg{B}$, respectively. 
Let $U_{\alpha}\in\Alg{A}$ and $U_{\beta}\in\Alg{B}$ be unitaries that leave the states invariant, i.e., 
\[
U_{\alpha}\alpha U_{\alpha}^{\dag}=\alpha
\quad\text{ and }\quad
U_{\beta}\beta U_{\beta}^{\dag}=\beta.
\]
Given a morphism $(\Alg{A},\alpha)\xrightarrow{\mE}(\Alg{B},\beta)$ in $\St$, the \define{Sutter--Tomamichel--Harrow (STH) rotated Petz recovery map} associated with the unitaries $U_{\alpha}$ and $U_{\beta}$ is given by 
\[
\retro^{\STH}_{\alpha,\mE}=\Ad_{U_{\alpha}^{\dag}}\circ\retro^{\Petz}_{\alpha,\mE}\circ\Ad_{U_{\beta}}
\equiv\Ad_{U_{\alpha}^{\dag}}\circ\Ad_{\alpha^{1/2}}\circ \mE^*\circ\Ad_{\beta^{-1/2}}\circ\Ad_{U_{\beta}}.
\]
\ed

The STH rotated Petz recovery map is a slight generalization of the version considered in~\cite{SuToHa16}.%
\footnote{In~\cite{SuToHa16}, these  maps were denoted $\mathcal{T}^{\varphi,\vartheta}_{\sigma,\mathcal{N}}$.}
 This special case considered in~\cite{SuToHa16} occurs when all the unitaries $U_{\alpha}$ and $U_{\beta}$ are taken to be diagonal with respect to eigenbases for $\alpha$ and $\beta$, respectively. The more general unitaries $U_{\alpha}$ and $U_{\beta}$ considered here are block-diagonal unitaries, where the blocks correspond to the eigenspaces of $\alpha$ and $\beta$, respectively. Note that if the dimensions of the eigenspaces are all $1$, then this automatically forces all the unitaries to be diagonal in these bases, so our above generalization allows one to consider unitaries that are not necessarily phases on the eigenspaces when the dimension of the eigenspaces are greater than $1$.

\bn
For each $(\Alg{A},\alpha)$ in $\St$, fix a unitary $U_{\alpha}\in\Alg{A}$ such that $U_{\alpha}\alpha U_{\alpha}^{\dag}=\alpha$. 
Then 
\[
\St\xrightarrow{\retro^{\STH}}\St^{\op}
\]
sending a morphism $(\Alg{A},\alpha)\xrightarrow{\mE}(\Alg{B},\beta)$ to $(\Alg{A},\alpha)\xleftarrow{\retro^{\STH}_{\alpha,\mE}}(\Alg{B},\beta)$ defines a retrodiction family and is a functor, i.e., it satisfies normalization and compositionality. 
\en

\bprf
First, since $\retro^{\STH}_{\alpha,\mE}$ is a composite of CPTP maps, it is CPTP. Second, it is state-preserving because $\Ad_{U_{\beta}}(\beta)=\beta$ and $\Ad_{U_{\alpha}^{\dag}}(\alpha)=\alpha$ by definition and because the Petz recovery map is state-preserving. 
This shows that $\retro^{\STH}$ is indeed well-defined and sends morphisms in $\St$ to morphisms in $\St$ in a contravariant manner. 
The normalization property follows from 
\[
\retro^{\STH}_{\alpha,\id_{\Alg{A}}}
=\Ad_{U_{\alpha}^{\dag}}\circ\retro^{\Petz}_{\alpha,\id_{\Alg{A}}}\circ\Ad_{U_{\alpha}}
=\Ad_{U_{\alpha}^{\dag}}\circ\id_{\Alg{A}}\circ\Ad_{U_{\alpha}}
=\id_{\Alg{A}}.
\]
Finally, given a pair $(\Alg{A},\alpha)\xrightarrow{\mE}(\Alg{B},\beta)\xrightarrow{\mF}(\Alg{C},\gamma)$ of composable morphisms in $\St$, 
\[
\begin{split}
\retro^{\STH}_{\alpha,\mE}\circ\retro^{\STH}_{\beta,\mF}
&=\Ad_{U_{\alpha}^{\dag}}\circ\retro^{\Petz}_{\alpha,\mE}\circ\Ad_{U_{\beta}}\circ\Ad_{U_{\beta}^{\dag}}\circ\retro^{\Petz}_{\beta,\mF}\circ\Ad_{U_{\gamma}}\\
&=\Ad_{U_{\alpha}^{\dag}}\circ\retro^{\Petz}_{\alpha,\mE}\circ\retro^{\Petz}_{\beta,\mF}\circ\Ad_{U_{\gamma}}\\
&=\Ad_{U_{\alpha}^{\dag}}\circ\retro^{\Petz}_{\alpha,\mF\circ\mE}\circ\Ad_{U_{\gamma}}\\
&=\retro^{\STH}_{\alpha,\mF\circ\mE},
\end{split}
\]
which proves compositionality of $\retro^{\STH}$. 
\eprf

\bn
\label{prop:Petzimpliesbayes}
When restricted to the subcategory $\CSt\subset\St$ of commutative $C^*$-algebras, all of the retrodiction families $\retro^{\Petz},\retro^{\Petz,t},\retro^{\Petz,\mu},$ and $\retro^{\STH}$ agree with standard (classical) Bayesian inversion, i.e., any of these retrodiction families applied to a morphism $(\C^{X},p)\xrightarrow{\mathcal{E}}(\C^{Y},q)$ gives $(\C^{X},p)\xleftarrow{\overline{\mathcal{E}}:=\retro^{\Bayes}_{p,\mathcal{E}}}(\C^{Y},q)$ from Proposition~\ref{prop:classicalBayes}.
\en

\bprf
This follows from the fact that all the unitaries in the formulas for these retrodiction maps cancel due to commutativity. Hence, all of these retrodiction families equal the Petz recovery retrodiction family on $\CSt$. The remaining part follows from the fact that the Petz recovery map gives the Bayesian inverse on commutative $C^*$-algebras, which is well-known~\cite{LiWi18}. 
\eprf

\section{The convexity of certain retrodiction families}

In general, the collection of retrodiction families, \emph{without} assuming compositionality/tensoriality properties, is a convex space. Namely, given any two retrodiction families $\retro$ and $\retro'$, the \define{convex combination} $\lambda\retro+(1-\lambda)\retro'$, determined by sending $(\Alg{A},\alpha)\xrightarrow{\mE}(\Alg{B},\beta)$ to 
\[
\left(\lambda\retro+(1-\lambda)\retro'\right)_{\alpha,\mE}:=\lambda\retro_{\alpha,\mE}+(1-\lambda)\retro'_{\alpha,\mE}\;,
\]
is a retrodiction family. In fact, if $\retro$ and $\retro'$ are normalizing, then so is $\lambda\retro+(1-\lambda)\retro'$. A closely related fact was already observed in Proposition~\ref{prop:RPmupreretro}. \emph{However}, the collection of retrodiction families that are compositional and/or tensorial is \emph{not} a convex space, nor should we expect it to be.
\footnote{It is not even obvious that the space of retrodiction families that are compositional and/or tensorial should be path-connected. Namely, why should there exist a continuous interpolation between two arbitrary retrodiction procedures, much less a linear path?}
In this section, we justify this with Proposition~\ref{prop:bitflipavgPetz} and Proposition~\ref{ex:JRSWWnotretro}, the latter of which shows that the JRSWW map is \emph{not} compositional \emph{nor} tensorial, thus clarifying some of the claims made in~\cite[Remark~2.4]{JRSWW18} 
(although it does satisfy a type of stabilization property, which will be discussed later). 

But before this, we make an observation that might suggest why convexity might not hold for retrodiction families that are compositional. We do this by analyzing the averaged rotated Petz recovery maps, which are (integrated) convex combinations of rotated Petz recovery maps. 

\bd
\label{defn:covariantchannel}
Let $(\Alg{A},\alpha)\xrightarrow{\mE}(\Alg{B},\beta)$ be a morphism in $\St$. The morphism $\mE$ is said to be \define{covariant}%
\footnote{It would be more precise to say that $\mE$ is \emph{time-symmetric covariant with respect to the modular Hamiltonians} associated with $\alpha$ and $\beta$. However, since this is the only covariance considered in this work, we use the shorthand ``covariant.'' This notion of covariance, its relation to Bayesian invertibility, and simple criteria for guaranteeing such covariance (including necessary modifications for non-faithful states) are discussed in~\cite{PaRuBayes,PaBayes,GPRR23}.}
iff 
$\Ad_{\beta^{it}}\circ\mE=\mE\circ\Ad_{\alpha^{it}}$
for all $t\in\R$.
\ed

\blem
\label{lem:covariantstabilizing}
Let $(\Alg{A},\alpha)\xrightarrow{\mE}(\Alg{B},\beta)\xrightarrow{\mF}(\Alg{C},\gamma)$ be a composable pair of morphisms in $\St$, and let $\mu$ and $\nu$ be two probability measures on $\R$. Then 
\[
\retro^{\Petz,\mu}_{\alpha,\mE}\circ\retro^{\Petz,\nu}_{\beta,\mF}
=\int_{-\infty}^{\infty}\int_{-\infty}^{\infty}\left(\Ad_{\alpha^{-is+1/2}}\circ \mE^*\circ\Ad_{\beta^{i(s-t)}}\circ\mF^*\circ\Ad_{\gamma^{it-1/2}}\right)d\mu(s)d\nu(t).
\]
Furthermore, if at least one of $\mF$ or $\mE$ is 
covariant (cf.\ Definition~\ref{defn:covariantchannel}),  
then 
\[
\retro^{\Petz,\mu}_{\alpha,\mE}\circ\retro^{\Petz,\nu}_{\beta,\mF}
=
\begin{cases}
\retro^{\Petz,\mu}_{\alpha,\mF\circ\mE}&\mbox{ if $\mF$ is covariant}\\
\retro^{\Petz,\nu}_{\alpha,\mF\circ\mE}&\mbox{ if $\mE$ is covariant.}
\end{cases}
\]
In particular, if both $\mE$ and $\mF$ are covariant, then $\retro^{\Petz,\mu}_{\alpha,\mE}\circ\retro^{\Petz,\nu}_{\beta,\mF}=\retro^{\Petz}_{\alpha,\mE}\circ\retro^{\Petz}_{\beta,\mF}=\retro^{\Petz}_{\alpha,\mF\circ\mE}$.
\elem

\bprf
Let $C\in\Alg{C}$ be arbitrary. Then
\[
\begin{split}
\left(\retro^{\Petz,\mu}_{\alpha,\mE}\circ\retro^{\Petz,\nu}_{\beta,\mF}\right)(C)&=\retro^{\Petz,\mu}_{\alpha,\mE}\left(\retro^{\Petz,\nu}_{\beta,\mF}(C)\right)\\
&=\retro^{\Petz,\mu}_{\alpha,\mE}\left(\int_{-\infty}^{\infty}\big(\Ad_{\beta^{-it+1/2}}\circ\mF^*\circ\Ad_{\gamma^{it-1/2}}\big)(C)d\nu(t)\right)\\
&=\int_{-\infty}^{\infty}\big(\retro^{\Petz,\mu}_{\alpha,\mE}\circ\Ad_{\beta^{-it+1/2}}\circ\mF^*\circ\Ad_{\gamma^{it-1/2}}\big)(C)d\nu(t)\\
&=\int_{-\infty}^{\infty}\left(\int_{-\infty}^{\infty}\big(\Ad_{\alpha^{-is+1/2}}\circ \mE^*\circ\Ad_{\beta^{i(s-t)}}\circ\mF^*\circ\Ad_{\gamma^{it-1/2}}\big)(C)d\mu(s)\right)d\nu(t).
\end{split}
\]
The third equality follows from linearity and continuity of $\retro^{\Petz,\mu}_{\alpha,\mE}$ together with the arithmetic properties of convergent integrals~\cite{Ab15,Ru76}. 

Now, suppose that $\mE$ is covariant. Then $\retro^{\Petz,t}_{\alpha,\mE}=\retro^{\Petz}_{\alpha,\mE}$ for all $t\in\R$ so that $\retro^{\Petz,\mu}_{\alpha,\mE}=\retro^{\Petz}_{\alpha,\mE}$ as well (since $\mu$ is a probability measure). Hence, 
\[
\begin{split}
\retro^{\Petz,\mu}_{\alpha,\mE}\circ\retro^{\Petz,\nu}_{\beta,\mF}
&=\retro^{\Petz}_{\alpha,\mE}\circ\retro^{\Petz,\nu}_{\beta,\mF}\\
&=\int_{-\infty}^{\infty}\left(\Ad_{\alpha^{1/2}}\circ \mE^*\circ\Ad_{\beta^{-it}}\circ\mF^*\circ\Ad_{\gamma^{it-1/2}}\right)d\nu(t)\\
&=\int_{-\infty}^{\infty}\left(\Ad_{\alpha^{1/2}}\circ\Ad_{\alpha^{-it}}\circ \mE^*\circ\mF^*\circ\Ad_{\gamma^{it-1/2}}\right)d\nu(t)\\
&=\int_{-\infty}^{\infty}\left(\Ad_{\alpha^{-it+1/2}}\circ (\mF\circ \mE)^*\circ\Ad_{\gamma^{it-1/2}}\right)d\nu(t)\\
&=\retro^{\Petz,\nu}_{\alpha,\mF\circ\mE}\;,
\end{split}
\]
where covariance of $\mE$ was used again in the third equality. 
A similar calculation shows that if $\mF$ is covariant, then 
\[
\retro^{\Petz,\mu}_{\alpha,\mE}\circ\retro^{\Petz,\nu}_{\beta,\mF}=\retro^{\Petz,\mu}_{\alpha,\mF\circ\mE}\;. \qedhere
\]
\eprf

\bc
\label{cor:compositionalforcovariant}
Let $(\Alg{A},\alpha)\xrightarrow{\mE}(\Alg{B},\beta)\xrightarrow{\mF}(\Alg{C},\gamma)$ be a composable pair of morphisms in $\St$ and let $\mu$ be a probability measure on $\R$. If at least one of $\mF$ or $\mE$ is 
covariant (cf.\ Definition~\ref{defn:covariantchannel}), 
then 
\[
\retro^{\Petz,\mu}_{\alpha,\mF\circ\mE}=\retro^{\Petz,\mu}_{\alpha,\mE}\circ\retro^{\Petz,\mu}_{\beta,\mF}\;,
\]
i.e., $\retro^{\Petz,\mu}$ is compositional on such a pair of composable morphisms. 
\ec

In fact, similar results hold for tensoriality. 

\blem
\label{lem:tensoravgPetz}
Let $(\Alg{A},\alpha)\xrightarrow{\mE}(\Alg{B},\beta)$ and $(\Alg{A}',\alpha')\xrightarrow{\mE'}(\Alg{B}',\beta')$ be two morphisms in $\St$, and let $\mu$ and $\mu'$ be two probability measures on $\R$. Then
\[
\retro^{\Petz,\mu}_{\alpha,\mE}\otimes\retro^{\Petz,\mu'}_{\alpha',\mE'}
=\int_{-\infty}^{\infty}\int_{-\infty}^{\infty}\left(\Ad_{\alpha^{-it+1/2}\otimes\alpha'^{-it'+1/2}}\right)\circ\left(\mE\otimes\mE'\right)^{*}\circ\left(\Ad_{\beta^{it-1/2}\otimes\beta'^{it'-1/2}}\right)d\mu(t)\,d\mu'(t').
\]
Furthermore, if at least one of $\mE'$ or $\mE$ is covariant (cf.\ Definition~\ref{defn:covariantchannel}),  
then 
\[
\retro^{\Petz,\mu}_{\alpha,\mE}\otimes\retro^{\Petz,\mu'}_{\alpha',\mE'}
=
\begin{cases}
\retro^{\Petz,\mu}_{\alpha\otimes\alpha',\mE\otimes\mE'} &\mbox{ if $\mE'$ is covariant}\\
\retro^{\Petz,\mu'}_{\alpha\otimes\alpha',\mE\otimes\mE'} &\mbox{ if $\mE$ is covariant.}\\
\end{cases}
\]
In particular, if both $\mE$ and $\mE'$ are covariant, then $\retro^{\Petz,\mu}_{\alpha,\mE}\otimes\retro^{\Petz,\mu'}_{\alpha',\mE'}
=\retro^{\Petz}_{\alpha,\mE}\otimes\retro^{\Petz}_{\alpha',\mE'}=\retro^{\Petz}_{\alpha\otimes\alpha',\mE\otimes\mE'}$.
\elem

\bprf
The first formula follows from Fubini's theorem, the interchange law for the tensor product and composition, as well as the properties of the adjoint action maps (partial details will be given momentarily in the proof of the second claim). To illustrate the second claim, first suppose that $\mE'$ is covariant. Then $\retro^{\Petz,\mu'}_{\alpha',\mE'}=\retro^{\Petz}_{\alpha',\mE'}$ due to covariance and hence
\[
\begin{split}
\retro^{\Petz,\mu}_{\alpha,\mE}\otimes\retro^{\Petz,\mu'}_{\alpha',\mE'}&=
\retro^{\Petz,\mu}_{\alpha,\mE}\otimes\retro^{\Petz}_{\alpha',\mE'}\\
&=\left(\int_{-\infty}^{\infty}\left(\Ad_{\alpha^{-it+1/2}}\circ \mE^*\circ\Ad_{\beta^{it-1/2}}\right)d\mu(t)\right)\otimes\left(\Ad_{\alpha'^{1/2}}\circ\mE'^*\circ\Ad_{\beta'^{-1/2}}\right)\\
&=\int_{-\infty}^{\infty}\left(\Ad_{\alpha^{-it+1/2}}\circ \mE^*\circ\Ad_{\beta^{it-1/2}}\right)\otimes\left(\Ad_{\alpha'^{1/2}}\circ \mE'^*\circ\Ad_{\beta'^{-1/2}}\right)d\mu(t)\\
&=\int_{-\infty}^{\infty}\left(\Ad_{\alpha^{-it+1/2}}\circ \mE^*\circ\Ad_{\beta^{it-1/2}}\right)\otimes\left(\Ad_{\alpha'^{-it+1/2}}\circ \mE'^*\circ\Ad_{\beta'^{it-1/2}}\right)d\mu(t)\\
&=\int_{-\infty}^{\infty}\left(\Ad_{\alpha^{-it+1/2}}\otimes\Ad_{\alpha'^{-it+1/2}}\right)\circ\left(\mE\otimes\mE'\right)^{*}\circ\left(\Ad_{\beta^{it-1/2}}\otimes\Ad_{\beta'^{it-1/2}}\right)d\mu(t)\\
&=\int_{-\infty}^{\infty}\Ad_{(\alpha\otimes\alpha')^{-it+1/2}}\circ\left(\mE\otimes\mE'\right)^*\circ\Ad_{(\beta\otimes\beta')^{it-1/2}}\,d\mu(t)\\
&=\retro^{\Petz,\mu}_{\alpha\otimes\alpha',\mE\otimes\mE'}\;.
\end{split}
\]
A similar calculation proves the claim for when $\mE$ is covariant. 
\eprf

\bc
\label{cor:tensorialforcovariant}
Let $(\Alg{A},\alpha)\xrightarrow{\mE}(\Alg{B},\beta)$ and $(\Alg{A}',\alpha')\xrightarrow{\mE'}(\Alg{B}',\beta')$ be two morphisms in $\St$, and let $\mu$ be a probability measure on $\R$. If at least one of $\mE'$ or $\mE$ is 
covariant (cf.\ Definition~\ref{defn:covariantchannel}), then
\[
\retro^{\Petz,\mu}_{\alpha,\mE}\otimes\retro^{\Petz,\mu}_{\alpha',\mE'}=\retro^{\Petz,\mu}_{\alpha\otimes\alpha',\mE\otimes\mE'}\;, 
\]
i.e., $\retro^{\Petz,\mu}$ is tensorial on this pair of morphisms. 
\ec

Therefore, if such covariance does not hold, one might suspect that the retrodiction family given by some averaged rotated Petz recovery map might be neither functorial nor tensorial in general. The following proposition provides justification for such suspicion. 

\bn
\label{prop:bitflipavgPetz}
Let $t,s\in\R$ be two distinct numbers and let $\lambda\in(0,1)$. Then 
$(1-\lambda)\retro^{\Petz,s}+\lambda\retro^{\Petz,t}$ is a retrodiction family that is normalized but not necessarily compositional nor tensorial. 
\en

\bprf
A stronger result is proved in Appendix~\ref{sec:bitflipavgPetz}. 
\eprf

Since convex combinations of compositional and tensorial retrodiction families need not be compositional nor tensorial, one might suspect that the JRSWW averaged rotated Petz recovery map might \emph{not} be compositional nor tensorial. 
Despite that suspicion, in~\cite[Remark~2.4]{JRSWW18}, it was claimed that the JRSSWW map satisfies some parallel and series composition rules. In what follows, we will clarify what composition rules are satisfied by first showing that the standard notion of parallel and series composition (i.e., tensoriality and compositionality) do not hold in general.

\bn
\label{ex:JRSWWnotretro}
The JRSWW averaged rotated Petz recovery map $\retro^{\JRSWW}$ is a normalized retrodiction family that is neither compositional nor tensorial. In particular, averaged rotated Petz recovery maps need not define compositional nor tensorial retrodiction families.
\en

\bprf
The proof is provided in Appendix~\ref{sec:JRSWW}. 
\eprf

Nevertheless, the JRSWW retrodiction family \emph{does} satisfy the stabilization property mentioned in~\cite[Remark~2.4]{JRSWW18}. 
In fact, we will now generalize the stabilization property and show that the one considered in~\cite[Remark~2.4]{JRSWW18} can be viewed as a consequence of the fact that
the identity map is covariant (cf.\ Corollary~\ref{cor:tensorialforcovariant} and Corollary~\ref{cor:Eotimesid}).
In other words, although averaged Petz recovery maps do not necessarily satisfy compositionality nor tensoriality on \emph{all} morphisms, they do \emph{when} one of the morphisms is covariant. We formally isolate this property in the following definition, motivated by the terminology from~\cite{JRSWW18}. 

\bd
Let $\retro:\St\to\St^{\op}$ be a retrodiction family.
\begin{enumerate}
\item
$\retro$ is \define{composition stabilizing}, or \define{$\circ$-stabilizing}, iff $\retro_{\alpha,\mE}\circ\retro_{\beta,\mF}=\retro_{\alpha,\mF\circ\mE}$ for all composable pairs $(\Alg{A},\alpha)\xrightarrow{\mE}(\Alg{B},\beta)\xrightarrow{\mF}(\Alg{C},\gamma)$ of morphisms in $\St$ at least one of which is covariant.
\item
$\retro$ is \define{tensor stabilizing}, or \define{$\otimes$-stabilizing}, iff $\retro_{\alpha,\mE}\otimes\retro_{\alpha',\mE'}=\retro_{\alpha\otimes\alpha',\mE\otimes\mE'}$ for all pairs $(\Alg{A},\alpha)\xrightarrow{\mE}(\Alg{B},\beta)$ and $(\Alg{A}',\alpha')\xrightarrow{\mE'}(\Alg{B}',\beta')$ of morphisms in $\St$ at least one of which is covariant. 
\end{enumerate} 
$\retro$ satisfies the \define{stabilization property} iff it is both composition and tensor stabilizing. 
\ed

\bc
\label{cor:Eotimesid}
Let $\retro$ be a normalizing retrodiction family that is tensor stabilizing. Then $\retro_{\alpha\otimes\alpha',\mE\otimes\id_{\Alg{A}'}}=\retro_{\alpha,\mE}\otimes\id_{\Alg{A}'}$ for all morphisms $(\Alg{A},\alpha)\xrightarrow{\mE}(\Alg{B},\beta)$ and objects $(\Alg{A}',\alpha')$ in $\St$. 
\ec

As mentioned earlier, Corollary~\ref{cor:Eotimesid} provides an alternative justification for the stabilization property mentioned in~\cite[Remark~2.4]{JRSWW18}. Therefore, our definition generalizes this property. 

\bc
\label{cor:avgrotatedPetzstable}
The following three facts regarding the stabilizing properties hold.
\begin{enumerate}[(a)]
\item
\label{item:avgstable}
Every averaged rotated Petz retrodiction family satisfies the stabilization property. 
\item
\label{item:monoidalstable}
Every retrodiction family that is compositional satisfies the compositional stabilization property. 
Similarly, every retrodiction family that is tensorial satisfies the tensorial stabilization property.
\item
\label{item:compstronger}
More generally, if a retrodiction family is normalizing, compositional, and $\otimes$-stabilizing,%
\footnote{In fact, if $\retro$ denotes the retrodiction family, it suffices to assume $\retro_{\alpha\otimes\gamma,\mE\otimes\id_{\Alg{C}}}=\retro_{\alpha,\mE}\otimes\retro_{\gamma,\id_{\Alg{C}}}$ and $\retro_{\gamma\otimes\alpha,\id_{\Alg{C}}\otimes\mE}=\retro_{\gamma,\id_{\Alg{C}}}\otimes\retro_{\alpha,\mE}$ for all morphisms $(\Alg{A},\alpha)\xrightarrow{\mE}(\Alg{B},\beta)$ and objects $(\Alg{C},\gamma)$ in $\St$, rather than the full $\otimes$-stabilizing property. The sufficiency of this follows from the proof. Also note that Corollary~\ref{cor:avgrotatedPetzstable}~(\ref{item:compstronger}) does not contradict Proposition~\ref{ex:JRSWWnotretro}, because $\retro^{\JRSWW}$ is not compositional.}
then the retrodiction family is necessarily tensorial.
\end{enumerate}
\ec

\bprf
Statement~(\ref{item:avgstable}) is exactly what Corollaries~\ref{cor:compositionalforcovariant} and~\ref{cor:tensorialforcovariant} say. 
Statement~(\ref{item:monoidalstable}) follows directly from the definitions. 
For statement~(\ref{item:compstronger}), let $\retro$ denote a retrodiction family that is normalizing, compositional, and $\otimes$-stabilizing. Let $(\Alg{A},\alpha)\xrightarrow{\mE}(\Alg{B},\beta)$ and $(\Alg{A}',\alpha')\xrightarrow{\mE'}(\Alg{B}',\beta')$ be two pairs of morphisms in $\St$. Then
\begin{align*}
\retro_{\alpha\otimes\alpha',\mE\otimes\mE'}
&=\retro_{\alpha\otimes\alpha',(\mE\otimes\id_{\Alg{B}'})\circ(\id_{\Alg{A}}\otimes\mE')} && \mbox{ by the interchange law}\\ 
&=\retro_{\alpha\otimes\alpha',\id_{\Alg{A}}\otimes\mE'}\circ\retro_{\alpha\otimes\beta',\mE\otimes\id_{\Alg{B}'}} && \mbox{ since $\retro$ is compositional}\\
&=\left(\retro_{\alpha,\id_{\Alg{A}}}\otimes\retro_{\alpha',\mE'}\right)\circ\left(\retro_{\alpha,\mE}\otimes\retro_{\beta',\id_{\Alg{B}'}}\right) && \mbox{ since $\retro$ is $\otimes$-stabilizing}\\
&=\left(\id_{\Alg{A}}\otimes\retro_{\alpha',\mE'}\right)\circ\left(\retro_{\alpha,\mE}\otimes\id_{\Alg{B}'}\right) && \mbox{ since $\retro$ is normalizing}\\
&=\left(\id_{\Alg{A}}\circ\retro_{\alpha,\mE}\right)\otimes\left(\retro_{\alpha',\mE'}\circ\id_{\Alg{B}'}\right) && \mbox{ by the interchange law}\\
&=\retro_{\alpha,\mE}\otimes\retro_{\alpha',\mE'} && \mbox{ by definition of the identity,}
\end{align*}
which proves the claim.
\eprf

Statement~(\ref{item:avgstable}) in Corollary~\ref{cor:avgrotatedPetzstable} (together with Proposition~\ref{ex:JRSWWnotretro}) provides a precise sense of what parallel and series composition properties averaged rotated Petz recovery maps satisfy. 
Statement~(\ref{item:monoidalstable}) stresses the facts that compositionality and tensoriality are stronger conditions than composition and tensor stabilization, respectively. Finally, statement~(\ref{item:compstronger}) says that compositionality with tensor stabilization together imply tensoriality, which suggests that compositionality is in some sense more fundamental than tensoriality.

\section{Inverting property for retrodiction families}

A natural property one might want of a retrodiction family is that the retrodiction of an invertible (reversible) process is the inverse (reverse) process, and is, in particular, independent of the prior. 

\bd
A retrodiction family $\St\xrightarrow{\retro}\St^{\op}$ is \define{inverting} iff 
$\retro_{\alpha,\mE}=\mE^{-1}$ whenever $(\Alg{A},\alpha)\xrightarrow{\mE}(\Alg{B},\beta)$ is an isomorphism in $\St$. 
\ed

\bn
\label{prop:Petzinverting}
Every rotated Petz retrodiction family is inverting. 
\en

Before proving this, we recall a useful fact, reformulated in algebraic and categorical terms. 

\blem
\label{lem:isoSt}
Isomorphisms in the category $\St$ are precisely state-preserving $*$-isomorphisms. 
\elem

A special case of this lemma is well-known in the quantum information literature. Namely, it says that a CPTP map from one matrix algebra to another has a CPTP inverse if and only if it is of the form $\Ad_{U}$ for some unitary $U$. In what follows, we provide a proof applicable to the more general algebraic setting.

\bprf
[Proof of Lemma~\ref{lem:isoSt}]
Let $(\Alg{A},\alpha)\xrightarrow{\mathcal{E}}(\Alg{B},\beta)$ be an isomorphism in $\St$. This means there exists a morphism  $(\Alg{A},\alpha)\xleftarrow{\mathcal{F}}(\Alg{B},\beta)$ in $\St$ such that $\mathcal{F}\circ\mathcal{E}=\id_{\Alg{A}}$ and $\mathcal{E}\circ\mathcal{F}=\id_{\Alg{B}}$.  
Write the Hilbert--Schmidt adjoints as $E:=\mathcal{E}^*$ and $F:=\mathcal{F}^*$, which are completely positive \emph{unital} maps. Then, for any $A\in\Alg{A}$, 
\[
E(A)^{\dag}E(A)
\le E(A^{\dag}A)
=E\big(F(B)^{\dag}F(B)\big)
\le E\big(F(B^{\dag}B)\big)
=B^{\dag}B
=E(A)^{\dag}E(A),
\]
where $B:=E(A)$ and where the Kadison--Schwarz inequality was used (the first for $E$ and the second for $F$). Since the outermost terms in this expression are equal, all intermediate terms are equal. Since this holds for arbitrary $A\in\Alg{A}$, the \emph{Multiplication Lemma} (cf.\ \cite[Lemma~4.3]{PaBayes} or \cite[Theorem~5]{Ma10}) then implies $E(A_{1})^{\dag}E(A_{2})=E(A_{1}^{\dag}A_{2})$ for all $A_{1},A_{2}\in\Alg{A}$, which means that $E$ is a $*$-homomorphism. Similarly, $F$ is a $*$-homomorphism. Since $E$ and $F$ are inverses of each other, this proves that $E$ is a $*$-isomorphism. 
Thus, $E^{*}=\mathcal{E}$ is a $*$-isomorphism. This shows that an isomorphism in $\St$ is a state-preserving $*$-isomorphism. 

Conversely, any state-preserving $*$-isomorphism is automatically a state-preserving CPTP map and therefore defines an isomorphism in $\St$. 
\eprf

Note that the assumption of complete positivity for morphisms in $\St$ is crucial in Lemma~\ref{lem:isoSt}. Indeed, although the transpose map on matrix algebras has itself as an inverse, it is not a $*$-homomorphism.

\bprf
[Proof of Proposition~\ref{prop:Petzinverting}]
Let $(\Alg{A},\alpha)\xrightarrow{\mE}(\Alg{B},\beta)$ be an isomorphism in $\St$, so that $\mE$ is a $*$-isomorphism by Lemma~\ref{lem:isoSt}. Since $\mE^{-1}$ is the Hilbert--Schmidt adjoint of $\mathcal{E}$ and because both $(\Alg{A},\alpha)\xrightarrow{\mE}(\Alg{B},\beta)$ and $(\Alg{B},\beta)\xrightarrow{\mE^{*}=\mE^{-1}}(\Alg{A},\alpha)$ are covariant~\cite{GPRR23}, we obtain 
\[
\begin{split}
\retro^{\Petz,t}_{\alpha,\mE}
&=\Ad_{\alpha^{1/2-it}}\circ\mE^*\circ\Ad_{\beta^{-1/2+it}}
=\Ad_{\alpha^{1/2}}\circ\mE^{-1}\circ\Ad_{\beta^{-1/2}}\\
&=\Ad_{\alpha^{1/2}}\circ\Ad_{\mE^{-1}(\beta^{-1/2})}\circ\mE^{-1}
=\Ad_{\alpha^{1/2}}\circ\Ad_{\alpha^{-1/2}}\circ\mE^{-1}
=\mE^{-1},
\end{split}
\]
where the third and fourth equalities used the fact that $\mE^{-1}$ is a $*$-homomorphism, and the fourth equality also used the functional calculus and state-preserving property to deduce  $\mE^{-1}(\beta^{-1/2})=\left(\mE^{-1}(\beta)\right)^{-1/2}=\alpha^{-1/2}$.  
\eprf

\bc
The averaged rotated Petz recovery retrodiction family $\retro^{\Petz,\mu}$ for any probability measure $\mu$ on $\R$ is inverting. 
\ec

These results suggest that perhaps normalizing retrodiction families that are compositional might satisfy a similar property. However, this is not the case, as the following example illustrates. 

\bx
In general, it is not the case that $\retro^{\STH}_{\alpha,\mE}=\mE^{-1}$ for an invertible morphism $(\Alg{A},\alpha)\xrightarrow{\mE}(\Alg{B},\beta)$ in $\St$. We illustrate this with an explicit counter-example, even in the case of when all unitary rotations are independent phases. Let $\Alg{A}=\matr_{2}(\C)=\Alg{B}$ and set $\alpha=\mathrm{diag}(p,1-p)$ for some $p\in[0,1]$. Let $\mE=\Ad_{\sigma_{x}}$, so that $\beta=\mathrm{diag}(1-p,p)$, and set
\[
U_{\alpha}=\begin{bmatrix}e^{i\theta}&0\\0&e^{i\phi}\end{bmatrix}
\quad\text{ and }\quad
U_{\beta}=\begin{bmatrix}e^{i\psi}&0\\0&e^{i\omega}\end{bmatrix}
\]
for some distinct $\theta,\phi,\psi,\omega\in[0,2\pi]$. Then 
\[
\retro^{\STH}_{\alpha,\mE}=\Ad_{U_{\alpha}^{\dag}}\circ\Ad_{\alpha^{1/2}}\circ\mE^{*}\circ\Ad_{\beta^{-1/2}}\circ\Ad_{U_{\beta}}=\Ad_{V}\circ\mE^{-1},
\]
where 
\[
V=\begin{bmatrix}e^{i(\omega-\theta)}&0\\0&e^{i(\psi-\phi)}\end{bmatrix}.
\]
This shows that $\retro^{\STH}_{\alpha,\mE}\ne\mE^{-1}$ even though $\retro^{\STH}$ satisfies universality, state-preservation, normalization, and compositionality properties. 
\ex

As another example of an inverting retrodiction family, we mention the recent proposal of Surace and Scandi for a state-retrieval map~\cite{SuSc22}. 

\bd
The \define{Surace--Scandi} retrodiction family $\retro^{\SS}:\St\to\St^{\op}$ sends each morphism $(\Alg{A},\alpha)\xrightarrow{\mE}(\Alg{B},\beta)$ to the unique morphism $(\Alg{B},\beta)\xrightarrow{\retro^{\SS}_{\alpha,\mE}}(\Alg{A},\alpha)$ satisfying the following conditions.
\begin{enumerate}[(a)]
\item
If $\mE$ is an isomorphism in $\St$, then $\retro^{\SS}_{\alpha,\mE}=\mE^{-1}$. 
\item
The map $\retro^{\SS}_{\alpha,\mE}\circ\mE$ satisfies the detailed balance condition 
\[
\left(\retro^{\SS}_{\alpha,\mE}\circ\mE\right)\circ\Ad_{\alpha^{1/2}}=\Ad_{\alpha^{1/2}}\circ\left(\retro^{\SS}_{\alpha,\mE}\circ\mE\right)^{*}
\]
with respect to the prior $\alpha$. 
\item
The eigenvalues of the linear map $\retro^{\SS}_{\alpha,\mE}\circ\mE:\Alg{A}\to\Alg{A}$ are all non-negative.
\item
The map $\retro^{\SS}_{\alpha,\mE}$ maximizes the determinant of the linear map $\retro^{\SS}_{\alpha,\mE}\circ\mE$ subject to all the previous constraints.
\end{enumerate}
\ed

From the very definition of the Surace--Scandi retrodiction family, it should seem unlikely that it is compositional. The intuitive reason is that the maximization step is `local' in the sense that it involves only a single morphism $(\Alg{A},\alpha)\xrightarrow{\mE}(\Alg{B},\beta)$, whereas the compositional property is a `global' property involving composable morphisms such as $(\Alg{A},\alpha)\xrightarrow{\mE}(\Alg{B},\beta)\xrightarrow{\mF}(\Alg{C},\gamma)$. There is no obvious reason that maximizing the determinant for the individual morphisms should compose together to form a morphism that maximizes the determinant for the composite. 

\bn
\label{prop:SSnotstabilizing}
The Surace--Scandi retrodiction family is normalizing and inverting, but it is not  compositional. In fact, it is not even composition stabilizing. 
\en

\bprf
The proof is provided in Appendix~\ref{sec:SSnotstabilizing}. 
\eprf

In the remainder of this section, we describe how the prior ``disappears'' for inverting retrodiction families. As it was pointed out in~\cite[Result~2]{AwBuSc21}, the Petz recovery map satisfies the property that it not only takes an invertible map to its inverse, but that this is the only possibility if the retrodiction does not depend on the prior. The following theorem is a generalization of this result to the setting of inverting retrodiction families. 

\bt
\label{thm:AwBuSc}
Let $\retro:\St\to\St^{\op}$ be an inverting retrodiction family, let $\Alg{A}$ and $\Alg{B}$ be $*$-isomorphic $C^*$-algebras, and let $\Alg{A}\xrightarrow{\mE}\Alg{B}$ be a CPTP map. Then the following are equivalent. 
\begin{enumerate}[I.]
\item
\label{item:Fiso}
The map $\mE$ is a $*$-isomorphism. 
\item
\label{item:RaFindep}
The morphism $\retro_{\alpha,\mE}:(\Alg{B},\mE(\alpha))\to(\Alg{A},\alpha)$ is independent of the (faithful) state $\alpha$ on $\Alg{A}$, in the sense that the underlying map $\retro_{\alpha,\mE}:\Alg{B}\to\Alg{A}$ is the same for all $\alpha$. 
\item
\label{item:RaFisFinv}
There exists a (faithful) state $\alpha$ on $\Alg{A}$ such that $\retro_{\alpha,\mE}=\mE^{-1}$. 
\end{enumerate}
\et

Although the following proof will follow almost the same line of reasoning as that of~\cite[Result~2]{AwBuSc21}, Theorem~\ref{thm:AwBuSc} is a generalization of~\cite[Result~2]{AwBuSc21} in at least two respects. First, our result holds for all finite-dimensional $C^*$-algebras. Second, since the Petz recovery map is just one example of an inverting retrodiction family by Proposition~\ref{prop:Petzinverting},~\cite[Result~2]{AwBuSc21} follows from Theorem~\ref{thm:AwBuSc}. Furthermore, our result proves that a similar property holds for all rotated Petz recovery maps, averaged rotated Petz recovery maps, and even the recovery map of Surace--Scandi, without the need for doing explicit computations involving the specific formulas used to define all these different retrodiction families. The only time an explicit computation was needed was in proving that these define inverting retrodiction families. 

\bprf[Proof of Theorem~\ref{thm:AwBuSc}]
{\color{white}{you found me!}}

\noindent
(\ref{item:Fiso}$\Rightarrow$\ref{item:RaFindep}) Suppose $\mE$ is a $*$-isomorphism. Then by definition of $\retro$ being inverting, $\retro_{\alpha,\mE}=\mE^{-1}$ for all states $\alpha$ on $\Alg{A}$. Hence,~\ref{item:RaFindep} holds (this also proves~\ref{item:RaFisFinv}).

\noindent
(\ref{item:RaFindep}$\Rightarrow$\ref{item:RaFisFinv}) Since $\retro_{\alpha,\mE}:\Alg{B}\to\Alg{A}$ is independent of $\alpha$, denote it by $\retro_{\mE}$. By the state-preserving assumption of a retrodiction family, $(\retro_{\mE}\circ\mE)(\alpha)=\alpha$ for all states $\alpha$ on $\Alg{A}$. 
Since all elements of $\Alg{A}$ are complex combinations of states, this implies $\retro_{\mE}\circ\mE=\id_{\Alg{A}}$. 
Since $\Alg{A}$ and $\Alg{B}$ are $*$-isomorphic and finite dimensional, this implies $\retro_{\mE}=\mE^{-1}$. Hence,~\ref{item:RaFisFinv} holds. 

\noindent
(\ref{item:RaFisFinv}$\Rightarrow$\ref{item:Fiso}) Let $\alpha$ be some faithful state on $\Alg{A}$ such that $\retro_{\alpha,\mE}=\mE^{-1}$. This says that both $\mE$ and $\mE^{-1}$ are CPTP maps between non-degenerate quantum probability spaces. Hence, by Lemma~\ref{lem:isoSt}, $\mE$ is a $*$-isomorphism. Hence,~\ref{item:Fiso} holds.
\eprf

\section{Involutive retrodiction}

An interesting property emerges when two averaged rotated Petz retrodiction families are composed in succession. The meaning of composing two retrodiction families in succession is to iterate retrodiction procedures (i.e., retrodicting a retrodiction). Intuitively, we might expect to get back something closely related to the \emph{original} process. 
This happens, for example, when taking the Hilbert--Schmidt adjoint of an arbitrary CPTP map. Indeed, in the case of closed system evolution described by a unitary $U$, the Hilbert--Schmidt adjoint is the reverse evolution described by $U^{\dag}$. Applying the adjoint again gives back $U$.
This is part of the reason why the Hilbert--Schmidt adjoint has often been viewed as a time-reversal assignment in the literature~\cite{Ba06,CGS17,CAZ21,SSGC22}.

From our perspective, however, the Hilbert--Schmidt adjoint is only reasonable in the setting where either one has unitary evolution (cf.\ Theorem~\ref{thm:AwBuSc}) or unital quantum channels. When one also has a prior in the form of a state that is not necessarily uniform, the Hilbert--Schmidt adjoint no longer takes the prediction back to the prior, nor is it necessarily a quantum channel (even upon rescaling)%
\footnote{An example is a discard-and-prepare channel whenever the prepared state is not proportional to the identity, i.e., it is not the maximally mixed state.}.
This is one of the reasons why we have introduced retrodiction families as assignments of the form $\St\to\St^{\op}$, as opposed to sending only quantum channels to quantum channels in the opposite direction without including the data of states. Nevertheless, we might hope a similar involutive property holds for at least some retrodiction families.
In this section, we show that among all averaged rotated Petz retrodiction maps, the original process is always obtained upon iteration only for the (un-rotated) Petz recovery map. In particular, the Petz recovery map bypasses the many no-go theorems in the literature regarding time-reversal symmetry.

\blem
\label{lem:iterateavgPetz}
Let $\mu$ and $\nu$ be two probability measures on $\R$ and let $(\Alg{A},\alpha)\xrightarrow{\mE}(\Alg{B},\beta)$ be a morphism in $\St$. Then%
\footnote{If $\retro,\retro':\St\to\St^{\op}$ are two retrodiction families, the meaning of $\retro\circ\retro'$, the composite of two retrodiction families, is technically $\retro^{\op}\circ\retro'$, where $\retro^{\op}:\St^{\op}\to\St$ is defined by the same formula as $\retro$ (which explains our abuse of notation).}
\[
\big(\retro^{\Petz,\nu}\circ\retro^{\Petz,\mu}\big)_{\alpha,\mE}:=\retro^{\Petz,\nu}_{\beta,\retro^{\Petz,\mu}_{\alpha,\mE}}=\int_{-\infty}^{\infty}\Ad_{\beta^{-ir}}\circ\mE\circ\Ad_{\alpha^{ir}}\,d(\mu\ast\nu)(r),
\]
where $\mu\ast\nu$ denotes the convolution of the measures $\mu$ and $\nu$. 
In particular, 
\[
\big(\retro^{\Petz,y}\circ\retro^{\Petz,x}\big)_{\alpha,\mE}=\Ad_{\beta^{-i(x+y)}}\circ\mE\circ\Ad_{\alpha^{i(x+y)}}
\]
for all $x,y\in\R$.
\elem

\bprf
By a direct computation, we obtain 
\[
\begin{split}
\retro^{\Petz,\nu}_{\beta,\retro^{\Petz,\mu}_{\alpha,\mE}}&=\int_{-\infty}^{\infty}\retro^{\Petz,t}_{\beta,\retro^{\Petz,\mu}_{\alpha,\mE}}\,d\nu(t)\\
&=\int_{-\infty}^{\infty}\Ad_{\beta^{-it+1/2}}\circ\left(\retro^{\Petz,\mu}_{\alpha,\mE}\right)^{*}\circ\Ad_{\alpha^{it-1/2}}\,d\nu(t)\\
&=\int_{-\infty}^{\infty}\Ad_{\beta^{-it+1/2}}\circ\left(\int_{-\infty}^{\infty}\Ad_{\alpha^{-is+1/2}}\circ\mE^*\circ\Ad_{\beta^{is-1/2}}\,d\mu(s)\right)^*\circ\Ad_{\alpha^{it-1/2}}\,d\nu(t)\\
&=\int_{-\infty}^{\infty}\int_{-\infty}^{\infty}\Ad_{\beta^{-i(s+t)}}\circ \mE\circ\Ad_{\alpha^{i(s+t)}}\,d\mu(s)\,d\nu(t).
\end{split}
\]
By changing variables $s=r-t$, this becomes 
\[
\retro^{\Petz,\nu}_{\beta,\retro^{\Petz,\mu}_{\alpha,\mE}}
=\int_{-\infty}^{\infty}\int_{-\infty}^{\infty}\Ad_{\beta^{-ir}}\circ \mE\circ\Ad_{\alpha^{ir}}\,d\mu(r-t)\,d\nu(t).
\]
Since the integrand now only depends on one of the two variables, namely $r$, Fubini's theorem implies the $t$ integral can be done first. The resulting inner integral is precisely the definition of the convolution (cf.\ \cite[Lemma~1.30]{Ka21}, for example), so the first claim follows.

The second claim follows by setting $\mu=\delta_{x}$ and $\nu=\delta_{y}$ to be the Dirac delta measures centered at $x$ and $y$, respectively, and then using the fact that $\delta_{x}\ast\delta_{y}=\delta_{x+y}$.
\eprf

\bd
Given a morphism $(\Alg{A},\alpha)\xrightarrow{\mE}(\Alg{B},\beta)$ in $\St$, a retrodiction family $\retro$ is said to be \define{involutive on $\mE$} iff
\[
(\retro\circ\retro)_{\alpha,\mE}:=\retro_{\beta,\retro_{\alpha,\mE}}=\mE. 
\]
A retrodiction family $\retro$ is \define{involutive} iff $\retro\circ\retro=\id_{\St}$.
\ed

\bd
A \define{retrodiction functor} is an involutive and inverting monoidal functor 
$\St\to\St^{\op}$
that acts as the identity on objects. 
\ed

\br
In particular, a retrodiction functor defines a dagger on the symmetric monoidal category $\St$, which turns $\St$ into a symmetric monoidal dagger category~\cite{AbCo04,Se07,Ba06,SSGC22}. 
Justifications for why a dagger structure is often used to model time-reversal symmetry is due to the many time-reversal-like properties it satisfies.
The difference between our work and~\cite{AbCo04,Se07,Ba06,SSGC22} is that our dagger structure is not on the category of quantum processes, but rather on the category of quantum \emph{states} and \emph{state-preserving} quantum processes. 

The formulation of retrodiction closest to ours that we are aware of is one that describes (classical) Bayesian inversion as a dagger structure on the category of states~\cite[Remark~13.10]{Fr20} (in fact, the result proved in~\cite{Fr20} is applicable to the setting of Markov categories). 
However, the explicit connection to retrodiction was not made in~\cite{Fr20}, nor was an axiomatic formulation emphasized as in our present work. This result was generalized to the quantum setting in~\cite[Remark~7.25]{PaBayes}, though that work only proved the claim for certain morphisms that were called \emph{Bayesian invertible} (this result also holds more generally for quantum Markov categories and implies~\cite[Remark~13.10]{Fr20}). Therefore, our work not only extends~\cite[Remark~13.10]{Fr20} and~\cite[Remark~7.25]{PaBayes} to all classical and quantum channels, but it also bypasses the many no-go theorems mentioned earlier, 
thus illustrating the importance of the prior in (inferential) time-reversal symmetry for quantum dynamics. 
\er

\bt
\label{thm:involutivePetz}
Among all the rotated Petz recovery retrodiction families $\{\retro^{\Petz,t}\}_{t\in\R}$, the only one that is involutive is $\retro^{\Petz}$, i.e., when $t=0$. 
\et

\bprf
Let $t\in\R$ and suppose $\retro^{\Petz,t}$ is involutive. Then, by Lemma~\ref{lem:iterateavgPetz}, 
\[
\retro^{\Petz,t}_{\beta,\retro^{\Petz,t}_{\alpha,\mE}}=\Ad_{\beta^{-2it}}\circ \mE\circ\Ad_{\alpha^{2it}}=\mE
\]
for every morphism $(\Alg{A},\alpha)\xrightarrow{\mE}(\Alg{B},\beta)$ in $\St$. This necessarily implies that $t=0$. To see this, consider the bit-flip channel from Appendix~\ref{sec:bitflipavgPetz}. Namely, $\Alg{A}=\matr_{2}(\C)=\Alg{B}$, $\mE=(1-p)\id+p\Ad_{\sigma_{x}}$, $\alpha=\left[\begin{smallmatrix}\theta&0\\0&1-\theta\end{smallmatrix}\right]$, and $\beta=\left[\begin{smallmatrix}\phi&0\\0&1-\phi\end{smallmatrix}\right]$. Temporarily set $\mE':=\Ad_{\beta^{-2it}}\circ \mE\circ\Ad_{\alpha^{2it}}$. Then, the Choi matrices $\mathscr{C}[\mE]$ and  $\mathscr{C}[\mE']$ of $\mE$ and $\mE'$, respectively, are 
\[
\mathscr{C}[\mE]=\begin{bmatrix}1-p&0&0&1-p\\0&p&p&0\\0&p&p&0\\1-p&0&0&1-p\end{bmatrix}
\]
and
\[
\mathscr{C}[\mE']=\begin{bmatrix}1-p&0&0&(1-p)\chi^{2it}\\0&p&p\omega^{-2it}&0\\0&p\omega^{2it}&p&0\\(1-p)\chi^{-2it}&0&0&1-p\end{bmatrix},
\]
where
\[
\chi:=\frac{\theta(1-\phi)}{(1-\theta)\phi}
\qquad\text{ and }\qquad
\omega:=\frac{(1-\theta)(1-\phi)}{\theta\phi}.
\]
In order for these two Choi matrices to be equal, it must be the case that 
\[
1=\chi^{2it}\equiv\cos\big(2\ln\left(\chi\right)t\big)+i\sin\big(2\ln\left(\chi\right)t\big)
\]
and
\[
1=\omega^{2it}\equiv\cos\big(2\ln\left(\omega\right)t\big)+i\sin\big(2\ln\left(\omega\right)t\big).
\]
In order for these equations to hold, the sine terms must vanish simultaneously, i.e., 
\[
0=\sin\big(2\ln\left(\chi\right)t\big)
\quad\text{ and }\quad
0=\sin\big(2\ln\left(\omega\right)t\big).
\]
If the logarithm terms $\ln\left(\chi\right)$ and $\ln\left(\omega\right)$ vanish, then the two Choi matrices are equal for all $t\in\R$. Hence, we need to choose $\theta$ and $p$ such that these logarithms do not vanish. This is easy enough, and it is sufficient to choose \emph{any} 
\[
\theta,p\in\left(0,\frac{1}{2}\right)\cup\left(\frac{1}{2},1\right).
\]
Once we do this, the logarithm terms necessarily do not vanish and the condition becomes that $t$ must be in the intersection of the two lattices generated by the zeros of these sine functions, i.e., 
\[
t\in\left(\frac{\pi}{\ln(\chi)}\Z\right)\bigcap\left(\frac{\pi}{\ln(\omega)}\Z\right)
\]
Now, we will choose $\theta$ and $p$ such that the ratio $\frac{\ln(\chi)}{\ln(\omega)}$ is irrational so that the only possible value of $t$ is $t=0$. Many such solutions exist. For example, take 
\[
\theta=\frac{1}{1+e^{\frac{\pi}{2}(\sqrt{2}+1)}}
\quad\text{ and }\quad
p=\frac{\sinh(\pi/2)}{\sinh(\pi/2)+\sinh(\pi/\sqrt{2})}
\qquad\text{(which implies } \phi=\frac{1}{1+e^{\frac{\pi}{2}(\sqrt{2}-1)}}\text{ ).}
\]
Then 
\[
\ln(\chi)=-\pi
\quad\text{ and }\quad
\ln(\omega)=\pi\sqrt{2}.
\]
Hence, with these values of $\theta$ and $p$, the only value of $t$ for which $\mE=\Ad_{\beta^{-2it}}\circ\mE\circ\Ad_{\alpha^{2it}}$ is $t=0$. 
\eprf

\br
One might object to Theorem~\ref{thm:involutivePetz} by claiming that it is unnatural to use $\retro^{\Petz,t}$ twice when performing an iterated retrodiction, and that one should instead use $\retro^{\Petz,-t}$ for the second instance, since then $\retro^{\Petz,-t}\circ\retro^{\Petz,t}=\id_{\St}$. However, this would require a \emph{higher-order} involution on the collection of all retrodiction families $\St\to\St^{\op}$ that takes some retrodiction family $\retro$ and constructs a new one $\retro^{\dag}$ such that in the special case of the rotated Petz recovery map one obtains $(\retro^{\Petz,t})^{\dag}=\retro^{\Petz,-t}$. With our definition of involutivity, we have imposed that one always uses the \emph{same} retrodiction family. Our definition agrees with the same type of involutivity axioms used in other works on time-reversal symmetry by implementing the usage of a dagger functor~\cite{AbCo04,Ba06,SSGC22}.
\er

\section{A summary of recovery maps and their properties}

In this work, we furnished axioms for retrodiction associated with priors and processes for arbitrary hybrid classical/quantum dynamics. We proved that the Petz recovery map provides an inferential time-reversal symmetry for such dynamics, bypassing many previous no-go theorems on time-reversal symmetry.
The defining properties of retrodiction relied on the language of monoidal categories and functors. Namely, retrodiction is a certain involutive monoidal functor on the category of states and processes that sends invertible morphisms to their inverses. We proved that among all rotated Petz recovery maps, averaged rotated Petz recovery maps, and the recovery map of Surace--Scandi, the only one that satisfies all our axioms of retrodiction is the original Petz recovery map. We have also shown specifically which properties fail by the other variants in the literature, including the universal recovery map of~\cite{JRSWW18}. 

To summarize these findings, we have created Table~\ref{tab:retroEx}, which indicates several of the examples of possible retrodiction assignments considered in this work together with the properties they satisfy. 
We have omitted the properties of universality, complete positivity, and state-preservation because all of the examples listed above satisfy these. 
We have also included ``Bayes on $\CSt$,'' which means that the retrodiction family equals Bayesian inversion when restricted to the subcategory of commutative $C^*$-algebras (justifications are provided in Proposition~\ref{prop:Petzimpliesbayes} and~\cite{LiWi18,PaBayes,PaRuBayes,GPRR23,SuSc22}).
As can be seen from this table, only the Petz recovery map satisfies all of our desiderata for retrodiction.

\begin{table}
\centering
\begin{tabular}{c | c c c c c c}
&$\retro^{\Petz}$ & $\retro^{\Petz,t}$ & $\retro^{\Petz,\mu}$ & $\retro^{\STH}$ &  $\retro^{\shriek}$ & $\retro^{\SS}$ \\
\hline
Normalization &\cmark & \cmark & \cmark & \cmark &\xmark & \cmark\\
$\circ$-stabilizing & \cmark & \cmark & \cmark & \cmark & \cmark & \xmark \\
Compositionality & \cmark & \cmark & \xmark & \cmark & \cmark & \xmark\\
$\otimes$-stabilizing & \cmark & \cmark & \cmark & \xmark & \cmark & ? \\
Tensoriality & \cmark & \cmark & \xmark & \xmark & \cmark & ?\\
Inverting & \cmark & \cmark & \cmark & \xmark & \xmark & \cmark\\
Involutivity & \cmark & \xmark & \xmark & \xmark & \xmark & \xmark \\
Bayes on $\CSt$ & \cmark & \cmark & \cmark & \cmark & \xmark & \xmark \\
\end{tabular}
\caption{This table indicates the properties satisfied of retrodiction families considered in this work. Note that $\retro^{\JRSWW}$ is an example of $\retro^{\Petz,\mu}$ and therefore also belongs in that column. We have not yet been able to determine whether $\retro^{\SS}$ is tensorial or $\otimes$-stabilizing. Regardless, the original Petz recovery map is the only one that satisfies all properties.}
\label{tab:retroEx}
\end{table}

We close our work with the statement of a remaining open problem, which we can now formulate as a precise mathematical problem to characterize retrodiction.

\bq
\label{que:retrocharacter}
What is the space of all retrodiction functors $\St\to\St^{\op}$? In particular, is $\retro^{\Petz}$, the Petz retrodiction functor, the \emph{unique} element in this space? If $\retro^{\Petz}$ is not the unique element, what are other examples of retrodiction, and what additional axioms are needed to characterize the Petz retrodiction functor? In particular, is \emph{classical} Bayesian inversion characterized as the unique involutive and inverting monoidal functor $\CSt\to\CSt^{\op}$?
\eq

\appendix

\section{The Hilbert--Schmidt inner product on \texorpdfstring{$C^*$}{C*}-algebras}
\label{sec:HSonfdC}

In the following, we provide a quick review of traces, the Hilbert--Schmidt inner product, the KMS inner product, and the Petz recovery map for finite-dimensional $C^*$-algebras. The reader is referred to~\cite[Chapter~8]{OhPe93} for more details. Given a finite-dimensional $C^*$-algebra, $\Alg{A}$, there is a canonical \define{trace}, $\tr$. Indeed, since every such $\Alg{A}$ is $*$-isomorphic to a direct sum of matrix algebras~\cite[Theorem~5.20 and Proposition~5.26]{Fa01}, say $\Phi:\Alg{A}\xrightarrow{\cong}\bigoplus_{x\in X}\matr_{m_{x}}(\C)$, the trace on $\Alg{A}$ is defined by 
\[
\tr(A):=\tr\big(\Phi(A)\big),
\]
where the latter trace is just the sum of the component-wise traces. This is independent of the choice of $*$-isomorphism $\Phi$ due to the cyclic property of the trace. Namely, if $\Psi$ is another such $*$-isomorphism, then $\Psi\circ\Phi^{-1}$ is a $*$-isomorphism of a direct sum of matrix algebras, which leaves the trace invariant. Hence,
\[
\tr\big(\Phi(A)\big)=\tr\Big(\big(\Psi\circ\Phi^{-1}\big)\big(\Phi(A)\big)\Big)=\tr\big(\Psi(A)\big). 
\]

Because of this, there is a canonical \define{Hilbert--Schmidt inner product} on $\Alg{A}$ as well. 
Namely, this inner product is defined on $(A_{1},A_{2})\in\Alg{A}\times\Alg{A}$ by 
\[
\<A_{1},A_{2}\>=\tr\big(A_{1}^{\dag}A_{2}\big)
\]
in terms of the above-defined trace. 

With such an inner product on every finite-dimensional $C^*$-algebra, if now $\Alg{A}\xrightarrow{\mE}\Alg{B}$ is a linear map, then $\mE^*$ is the \define{Hilbert--Schmidt adjoint} of $\mE$ and is defined as the adjoint of $\mE$ with respect to the Hilbert--Schmidt inner products on $\Alg{B}$ and $\Alg{A}$. Namely, $\Alg{A}\xleftarrow{\mE^*}\Alg{B}$ is the unique linear map satisfying
\[
\tr\big(\mE^*(B)^{\dag}A\big)\equiv\<\mE^*(B),A\>=\<B,\mE(A)\>\equiv\tr\big(B^{\dag}\mE(A)\big)
\]
for all $A\in\Alg{A}$ and $B\in\Alg{B}$. 

Given a state-preserving morphism $(\Alg{A},\alpha)\xrightarrow{\mE}(\Alg{B},\beta)$ of non-degenerate quantum probability spaces, the \define{Petz recovery map} $\retro^{\Petz}_{\alpha,\mE}$ is defined as the adjoint, not with respect to the Hilbert--Schmidt inner product, but with respect to the following inner products on $\Alg{A}$ and $\Alg{B}$. The \define{KMS inner product} on $\Alg{A}$ associated with the faithful state $\alpha$ is defined by sending $(A_{1},A_{2})\in\Alg{A}\times\Alg{A}$ to 
\[
\<A_{1},A_{2}\>_{\alpha}:=\tr\Big(A_{1}^{\dag}\alpha^{-1/2}A_{2}\alpha^{-1/2}\Big).
\]
The Petz recovery map $\retro^{\Petz}_{\alpha,\mE}$ is therefore the unique map that satisfies
\[
\big\<B,\mE(A)\big\>_{\beta}=\left\<\retro^{\Petz}_{\alpha,\mE}(B),A\right\>_{\alpha}
\]
for all $A\in\Alg{A}$ and $B\in\Alg{B}$. Writing this out explicitly gives
\[
\tr\Big(B^{\dag}\beta^{-1/2}\mE(A)\beta^{-1/2}\Big)
=\tr\Big(\retro^{\Petz}_{\alpha,\mE}(B)^{\dag}\alpha^{-1/2}A\alpha^{-1/2}\Big).
\]
From this expression, the formula for the Petz recovery map in Definition~\ref{defn:Petzmaps} follows.

\section{Convex sums of rotated Petz recovery maps}
\label{sec:bitflipavgPetz}

In this appendix, we prove Proposition~\ref{prop:bitflipavgPetz}. In fact, we even show that it suffices to work with the class of bit-flip channels on qubits to find counter-examples to compositionality and tensoriality, even in the symmetric case when $\lambda=\frac{1}{2}$ and $s=-t$ in Proposition~\ref{prop:bitflipavgPetz}. Since it is known that the bit flip channel is covariant only with respect to the uniform state~\cite{PaRuBayes}, we need to provide a counter-example where at least two of the states are non-uniform states. 

Let $\Alg{C}=\Alg{B}=\Alg{A}=\matr_{2}(\C)$ denote the $C^*$-algebra of $2\times2$ matrices. Let $\alpha=\left[\begin{smallmatrix}\theta&0\\0&1-\theta\end{smallmatrix}\right]$, $\beta=\left[\begin{smallmatrix}\phi&0\\0&1-\phi\end{smallmatrix}\right]$ and $\gamma=\left[\begin{smallmatrix}\psi&0\\0&1-\psi\end{smallmatrix}\right]$ be states on $\Alg{A}$, $\Alg{B}$, and $\Alg{C}$, respectively, for some $\theta,\phi,\psi\in(0,1)$. Let $\Omega_{p}:\matr_{2}(\C)\to\matr_{2}(\C)$ be the bit-flip channel weighted by $p\in[0,1]$ and given explicitly by 
\[
\Omega_{p}=(1-p)\id+p\Ad_{\sigma_{x}}, \qquad\text{ where } \sigma_{x}=\begin{bmatrix}0&1\\1&0\end{bmatrix}.
\]
Note the set of bit-flip channels is closed under composition, and in fact
\[
\Omega_{q}\circ\Omega_{p}=\Omega_{(1-q)p+q(1-p)}
\]
for all $p,q\in[0,1]$. 
Now, set  $\Alg{A}\xrightarrow{\mE}\Alg{B}$ and $\Alg{B}\xrightarrow{\mF}\Alg{C}$ to be the bit-flip channels $\Omega_{p}$ and $\Omega_{q}$, respectively, where $p,q\in[0,1]$ are arbitrary for now. Then $(\Alg{A},\alpha)\xrightarrow{\mE}(\Alg{B},\beta)\xrightarrow{\mF}(\Alg{C},\gamma)$ is a composable pair of morphisms in $\St$ if and only if 
\[
\phi=(1-p)\theta+p(1-\theta)
\quad\text{ and }\quad
\psi=(1-q)\phi+q(1-\phi).
\]
We will therefore assume these conditions are always met in this example. 
The associated rotated Petz recovery maps are given by 
\[
\retro^{\Petz,t}_{\alpha,\mE}=(1-p)\Ad_{\left[\begin{smallmatrix}\phi^{it-1/2}\theta^{-it+1/2}&0\\0&(1-\phi)^{it-1/2}(1-\theta)^{-it+1/2}\end{smallmatrix}\right]}
+p\Ad_{\left[\begin{smallmatrix}0&(1-\phi)^{it-1/2}\theta^{-it+1/2}\\\phi^{it-1/2}(1-\theta)^{-it+1/2}&0\end{smallmatrix}\right]}
\]
and similarly
\[
\retro^{\Petz,t}_{\beta,\mF}=(1-q)\Ad_{\left[\begin{smallmatrix}\psi^{it-1/2}\phi^{-it+1/2}&0\\0&(1-\psi)^{it-1/2}(1-\phi)^{-it+1/2}\end{smallmatrix}\right]}
+q\Ad_{\left[\begin{smallmatrix}0&(1-\psi)^{it-1/2}\phi^{-it+1/2}\\\psi^{it-1/2}(1-\phi)^{-it+1/2}&0\end{smallmatrix}\right]}\;.
\]
Several simplifications occur when $q=\frac{1}{2}$, so we will henceforth assume this. In particular, $\gamma=\frac{1}{2}\mathds{1}_{2}$, i.e., $\psi=\frac{1}{2}$, so that 
\[
\retro^{\Petz,t}_{\beta,\mF}=\Ad_{\left[\begin{smallmatrix}\phi^{-it+1/2}&0\\0&(1-\phi)^{-it+1/2}\end{smallmatrix}\right]}
+\Ad_{\left[\begin{smallmatrix}0&\phi^{-it+1/2}\\(1-\phi)^{-it+1/2}&0\end{smallmatrix}\right]}\;.
\]
In addition, since $\mF\circ\mE=\Omega_{1/2}$ as a function, we obtain
\[
\retro^{\Petz,t}_{\alpha,\mF\circ\mE}=\Ad_{\left[\begin{smallmatrix}\theta^{-it+1/2}&0\\0&(1-\theta)^{-it+1/2}\end{smallmatrix}\right]}
+\Ad_{\left[\begin{smallmatrix}0&\theta^{-it+1/2}\\(1-\theta)^{-it+1/2}&0\end{smallmatrix}\right]}\;.
\]
Now, let $\retro:=\lambda\retro^{\Petz,t}+(1-\lambda)\retro^{\Petz,s}$ be the convex combination of two rotated Petz recovery maps with $t,s\in\R$ weighted by some $\lambda\in(0,1)$. For our purposes, it suffices to consider the symmetric combination given by $s=-t$ and $\lambda=\frac{1}{2}.$ 
Then
\[
\begin{split}
\retro_{\alpha,\mE}&=\frac{1-p}{2}\Ad_{\left[\begin{smallmatrix}\phi^{it-1/2}\theta^{-it+1/2}&0\\0&(1-\phi)^{it-1/2}(1-\theta)^{-it+1/2}\end{smallmatrix}\right]}
+\frac{p}{2}\Ad_{\left[\begin{smallmatrix}0&(1-\phi)^{it-1/2}\theta^{-it+1/2}\\\phi^{it-1/2}(1-\theta)^{-it+1/2}&0\end{smallmatrix}\right]}\\
&+\frac{1-p}{2}\Ad_{\left[\begin{smallmatrix}\phi^{-it-1/2}\theta^{it+1/2}&0\\0&(1-\phi)^{-it-1/2}(1-\theta)^{it+1/2}\end{smallmatrix}\right]}
+\frac{p}{2}\Ad_{\left[\begin{smallmatrix}0&(1-\phi)^{-it-1/2}\theta^{it+1/2}\\\phi^{-it-1/2}(1-\theta)^{it+1/2}&0\end{smallmatrix}\right]},
\end{split}
\]
\[
\begin{split}
\retro_{\beta,\mF}&=\frac{1}{2}\Ad_{\left[\begin{smallmatrix}\phi^{-it+1/2}&0\\0&(1-\phi)^{-it+1/2}\end{smallmatrix}\right]}
+\frac{1}{2}\Ad_{\left[\begin{smallmatrix}0&\phi^{-it+1/2}\\(1-\phi)^{-it+1/2}&0\end{smallmatrix}\right]}\\
&+\frac{1}{2}\Ad_{\left[\begin{smallmatrix}\phi^{it+1/2}&0\\0&(1-\phi)^{it+1/2}\end{smallmatrix}\right]}
+\frac{1}{2}\Ad_{\left[\begin{smallmatrix}0&\phi^{it+1/2}\\(1-\phi)^{it+1/2}&0\end{smallmatrix}\right]}, 
\end{split}
\]
and 
\[
\begin{split}
\retro_{\alpha,\mF\circ\mE}&=\frac{1}{2}\Ad_{\left[\begin{smallmatrix}\theta^{-it+1/2}&0\\0&(1-\theta)^{-it+1/2}\end{smallmatrix}\right]}
+\frac{1}{2}\Ad_{\left[\begin{smallmatrix}0&\theta^{-it+1/2}\\(1-\theta)^{-it+1/2}&0\end{smallmatrix}\right]}\\
&+\frac{1}{2}\Ad_{\left[\begin{smallmatrix}\theta^{it+1/2}&0\\0&(1-\theta)^{it+1/2}\end{smallmatrix}\right]}
+\frac{1}{2}\Ad_{\left[\begin{smallmatrix}0&\theta^{it+1/2}\\(1-\theta)^{it+1/2}&0\end{smallmatrix}\right]},
\end{split}
\]
the last of which is the retrodiction map associated with the composite. 
However, the composite of the individual retrodiction maps is given by 
\[
\begin{split}
\retro_{\alpha,\mE}\circ\retro_{\beta,\mF}&=\frac{1}{4}\left(\retro^{\Petz,t}_{\alpha,\mE}\circ\retro^{\Petz,t}_{\beta,\mF}+\retro^{\Petz,-t}_{\alpha,\mE}\circ\retro^{\Petz,-t}_{\beta,\mF}+\retro^{\Petz,-t}_{\alpha,\mE}\circ\retro^{\Petz,t}_{\beta,\mF}+\retro^{\Petz,t}_{\alpha,\mE}\circ\retro^{\Petz,-t}_{\beta,\mF}\right)\\
&=\frac{1}{2}\retro_{\alpha,\mF\circ\mE}\\
&+\frac{1-p}{4}\Ad_{\left[\begin{smallmatrix}\phi^{-2it}\theta^{it+1/2}&0\\0&(1-\phi)^{-2it}(1-\theta)^{it+1/2}\end{smallmatrix}\right]}
+\frac{p}{4}\Ad_{\left[\begin{smallmatrix}0&(1-\phi)^{-2it}\theta^{it+1/2}\\\phi^{-2it}(1-\theta)^{it+1/2}&0\end{smallmatrix}\right]}\\
&+\frac{1-p}{4}\Ad_{\left[\begin{smallmatrix}0&\phi^{-2it}\theta^{it+1/2}\\(1-\phi)^{-2it}(1-\theta)^{it+1/2}&0\end{smallmatrix}\right]}
+\frac{p}{4}\Ad_{\left[\begin{smallmatrix}(1-\phi)^{-2it}\theta^{it+1/2}&0\\0&\phi^{-2it}(1-\theta)^{it+1/2}\end{smallmatrix}\right]}\\
&+\frac{1-p}{4}\Ad_{\left[\begin{smallmatrix}\phi^{2it}\theta^{-it+1/2}&0\\0&(1-\phi)^{2it}(1-\theta)^{-it+1/2}\end{smallmatrix}\right]}
+\frac{p}{4}\Ad_{\left[\begin{smallmatrix}0&(1-\phi)^{2it}\theta^{-it+1/2}\\\phi^{2it}(1-\theta)^{-it+1/2}&0\end{smallmatrix}\right]}\\
&+\frac{1-p}{4}\Ad_{\left[\begin{smallmatrix}0&\phi^{2it}\theta^{-it+1/2}\\(1-\phi)^{2it}(1-\theta)^{-it+1/2}&0\end{smallmatrix}\right]}+\frac{p}{4}\Ad_{\left[\begin{smallmatrix}(1-\phi)^{2it}\theta^{-it+1/2}&0\\0&\phi^{2it}(1-\theta)^{-it+1/2}\end{smallmatrix}\right]}.
\end{split}
\]
This suggests that $\retro_{\alpha,\mE}\circ\retro_{\beta,\mF}\ne\retro_{\alpha,\mF\circ\mE}$, but it may be the case that surprising cancellations occur. To be sure, let us plug in a non-diagonal matrix such as $\left[\begin{smallmatrix}0&1\\0&0\end{smallmatrix}\right]$. This gives
\[
\retro_{\alpha,\mF\circ\mE}\left(\begin{bmatrix}0&1\\0&0\end{bmatrix}\right)=\sqrt{\theta(1-\theta)}\cos\left(\ln\left(\frac{\theta}{1-\theta}\right)t\right)\sigma_{x}
\]
and
\[
\begin{split}
\left(\retro_{\alpha,\mE}\circ\retro_{\beta,\mF}\right)\left(\begin{bmatrix}0&1\\0&0\end{bmatrix}\right)
=\frac{\sqrt{\theta(1-\theta)}}{2}\Bigg[&\cos\left(\ln\left(\frac{\theta}{1-\theta}\right)t\right)+p\cos\left(\ln\left(\frac{\theta\phi^{2}}{(1-\theta)(1-\phi)^{2}}\right)t\right)+\\
&+(1-p)\cos\left(\ln\left(\frac{\theta(1-\phi)^2}{(1-\theta)\phi^2}\right)t\right)\Bigg]\sigma_{x}.
\end{split}
\]
At this point, it may already be clear that these two expressions are different, but just to be very concrete, let us consider the special case where
\[
\theta=\frac{e^{2\pi}}{1+e^{2\pi}}\;,
\qquad
p=\frac{\tanh\left(\frac{\pi}{2}\right)}{2\sinh(\pi)}
\qquad\text{(which implies } \phi=\frac{e^{\pi}}{1+e^{\pi}}\text{ ),}
\quad\text{ and }\quad
t=\frac{1}{2}.
\]
Then 
\[
\retro_{\alpha,\mF\circ\mE}\left(\begin{bmatrix}0&1\\0&0\end{bmatrix}\right)
=\left(\frac{-1}{2\cosh(\pi)}\right)\sigma_{x}\;,
\quad\text{ while }\quad
\left(\retro_{\alpha,\mE}\circ\retro_{\beta,\mF}\right)\left(\begin{bmatrix}0&1\\0&0\end{bmatrix}\right)=0.
\]
This illustrates that $\retro_{\alpha,\mE}\circ\retro_{\beta,\mF}\ne\retro_{\alpha,\mF\circ\mE}$, i.e., compositionality fails in general (in fact, for most values of $t$). 

We will also use this example to show that tensoriality fails. In this regard, consider another morphism $(\Alg{A}',\alpha')\xrightarrow{\mE'}(\Alg{B}',\beta')$, where $\Alg{B}'=\matr_{2}(\C)=\Alg{A}'$, $\mE'=\Omega_{p'}$, and $\alpha'$ and $\beta'$ are given by the density matrices 
\[
\alpha'=\begin{bmatrix}\theta'&0\\0&1-\theta'\end{bmatrix}
\quad\text{ and }\quad
\beta'=\begin{bmatrix}\phi'&0\\0&1-\phi'\end{bmatrix}.
\]
It suffices to set $p=\frac{1}{2}=p'$. In this case, 
plugging in the matrix $\left[\begin{smallmatrix}0&1\\0&0\end{smallmatrix}\right]\otimes\left[\begin{smallmatrix}0&1\\0&0\end{smallmatrix}\right]$ into 
$\retro^{\Petz,t}_{\alpha,\mE}\otimes\retro^{\Petz,t'}_{\alpha',\mE'}$
gives
\[
\left(\retro^{\Petz,t}_{\alpha,\mE}\otimes\retro^{\Petz,t'}_{\alpha',\mE'}\right)\left(\begin{bmatrix}0&1\\0&0\end{bmatrix}\otimes\begin{bmatrix}0&1\\0&0\end{bmatrix}\right)
=
\sqrt{\theta\theta^{\perp}\theta'\theta'^{\perp}}
\begin{bmatrix}
0&\left(\frac{\theta^{\perp}}{\theta}\right)^{it}\\
\left(\frac{\theta^{\perp}}{\theta}\right)^{-it}&0
\end{bmatrix}
\otimes
\begin{bmatrix}
0&\left(\frac{\theta'^{\perp}}{\theta'}\right)^{it'}\\
\left(\frac{\theta'^{\perp}}{\theta'}\right)^{-it'}&0
\end{bmatrix}
,
\]
where we temporarily introduced the notation $r^{\perp}:=1-r$ for $r\in\R$. 
Therefore, 
\[
\left(\retro_{\alpha,\mE}\otimes\retro_{\alpha',\mE'}\right)\left(\begin{bmatrix}0&1\\0&0\end{bmatrix}\otimes\begin{bmatrix}0&1\\0&0\end{bmatrix}\right)
=\sqrt{\theta\theta^{\perp}\theta'\theta'^{\perp}}\cos\left(\ln\left(\frac{\theta^{\perp}}{\theta}\right)t\right)\cos\left(\ln\left(\frac{\theta'^{\perp}}{\theta'}\right)t\right)\sigma_{x}\otimes\sigma_{x}\;,
\]
while $\retro_{\alpha\otimes\alpha',\mE\otimes\mE'}=\frac{1}{2}\retro^{\Petz,t}_{\alpha\otimes\alpha',\mE\otimes\mE'}+\frac{1}{2}\retro^{\Petz,-t}_{\alpha\otimes\alpha',\mE\otimes\mE'}$ gives 
\[
\begin{split}
\retro_{\alpha\otimes\alpha',\mE\otimes\mE'}\left(\begin{bmatrix}0&1\\0&0\end{bmatrix}\otimes\begin{bmatrix}0&1\\0&0\end{bmatrix}\right)
=\sqrt{\theta\theta^{\perp}\theta'\theta'^{\perp}}&\bigg[\cos\left(\ln\left(\frac{\theta^{\perp}}{\theta}\right)t\right)\cos\left(\ln\left(\frac{\theta'^{\perp}}{\theta'}\right)t\right)\sigma_{x}\otimes\sigma_{x}\\
&+\sin\left(\ln\left(\frac{\theta^{\perp}}{\theta}\right)t\right)\sin\left(\ln\left(\frac{\theta'^{\perp}}{\theta'}\right)t\right)\sigma_{y}\otimes\sigma_{y}\bigg].
\end{split}
\]
These two results are manifestly different since for most values of $\theta$ and $\theta'$ the term involving $\sigma_{y}\otimes\sigma_{y}$ need not vanish. 
Hence, $\retro_{\alpha,\mE}\otimes\retro_{\alpha',\mE'}\ne\retro_{\alpha\otimes\alpha',\mE\otimes\mE'}$, i.e., $\retro$ is not tensorial.

\section{The JRSWW retrodiction family is neither compositional nor tensorial}
\label{sec:JRSWW}

In this appendix, we prove Proposition~\ref{ex:JRSWWnotretro}. As in Appendix~\ref{sec:bitflipavgPetz}, we show that it suffices to work with the class of bit-flip channels on qubits to find counter-examples to compositionality and tensoriality.
The same composable pair $(\Alg{A},\alpha)\xrightarrow{\mE}(\Alg{B},\beta)\xrightarrow{\mF}(\Alg{C},\gamma)$ as in Appendix~\ref{sec:bitflipavgPetz} will provide such a counter-example. In particular, we will again assume $q=\frac{1}{2}$ and keep $\theta$ and $\phi$ arbitrary for the moment (the fact that $\psi=\frac{1}{2}$ follows from $q=\frac{1}{2}$). First, 
\[
\retro^{\JRSWW}_{\alpha,\mF\circ\mE}=\int_{-\infty}^{\infty}\left(\frac{\pi}{\cosh(2\pi t)+1}\right)\left(\Ad_{\left[\begin{smallmatrix}\theta^{-it+1/2}&0\\0&(1-\theta)^{-it+1/2}\end{smallmatrix}\right]}+\Ad_{\left[\begin{smallmatrix}0&\theta^{-it+1/2}\\(1-\theta)^{-it+1/2}&0\end{smallmatrix}\right]}\right)dt.
\]
Meanwhile, by Lemma~\ref{lem:covariantstabilizing} or direct calculation, 
\[
\begin{split}
\retro^{\JRSWW}_{\alpha,\mE}\circ\retro^{\JRSWW}_{\beta,\mF}=
\int_{-\infty}^{\infty}\int_{-\infty}^{\infty}d\mu(t)d\mu(s)&\Bigg((1-p)\Ad_{\left[\begin{smallmatrix}\phi^{i(t-s)}\theta^{-it+1/2}&0\\0&(1-\phi)^{i(t-s)}(1-\theta)^{-it+1/2}\end{smallmatrix}\right]}\\
&+(1-p)\Ad_{\left[\begin{smallmatrix}0&\phi^{i(t-s)}\theta^{-it+1/2}\\(1-\phi)^{i(t-s)}(1-\theta)^{-it+1/2}&0\end{smallmatrix}\right]}\\
&+p\Ad_{\left[\begin{smallmatrix}0&(1-\phi)^{i(t-s)}\theta^{-it+1/2}\\\phi^{i(t-s)}(1-\theta)^{-it+1/2}&0\end{smallmatrix}\right]}\\
&+p\Ad_{\left[\begin{smallmatrix}(1-\phi)^{i(t-s)}\theta^{-it+1/2}&0\\0&\phi^{i(t-s)}(1-\theta)^{-it+1/2}\end{smallmatrix}\right]}\Bigg),
\end{split}
\]
where $\mu$ is the probability measure from Example~\ref{ex:JRSWW}. 
Plugging in the matrix $\left[\begin{smallmatrix}0&1\\0&0\end{smallmatrix}\right]$ into the first option gives
\[
\begin{split}
\retro^{\JRSWW}_{\alpha,\mF\circ\mE}\left(\begin{bmatrix}0&1\\0&0\end{bmatrix}\right)
&=\int_{-\infty}^{\infty}\left(\frac{\pi\sqrt{\theta(1-\theta)}}{\cosh(2\pi t)+1}\right)
\begin{bmatrix}
0&\left(\frac{\theta^{\perp}}{\theta}\right)^{it}\\
\left(\frac{\theta^{\perp}}{\theta}\right)^{-it}&0
\end{bmatrix}
dt\\
&=\int_{0}^{\infty}\left(\frac{2\pi\sqrt{\theta(1-\theta)}}{\cosh(2\pi t)+1}\right)\cos\left(\ln\left(\frac{\theta^{\perp}}{\theta}\right)t\right)dt\,\sigma_{x},
\end{split}
\]
where $r^{\perp}:=1-r$ as in Appendix~\ref{sec:bitflipavgPetz}. 
Plugging that same matrix into the second option gives
\[
\begin{split}
\Big(\retro^{\JRSWW}_{\beta,\mF}\circ&\retro^{\JRSWW}_{\alpha,\mE}\Big)\left(\begin{bmatrix}0&1\\0&0\end{bmatrix}\right)
=\int_{-\infty}^{\infty}\int_{-\infty}^{\infty}\left(\frac{\pi^2\sqrt{\theta(1-\theta)}}{[\cosh(2\pi t)+1][\cosh(2\pi s)+1]}\right)\\
&\times\begin{bmatrix}0&(1-p)\chi^{-it}\left(\frac{\phi^{\perp}}{\phi}\right)^{is}+p\omega^{it}\left(\frac{\phi^{\perp}}{\phi}\right)^{-is}\\(1-p)\chi^{it}\left(\frac{\phi^{\perp}}{\phi}\right)^{-is}+p\omega^{-it}\left(\frac{\phi^{\perp}}{\phi}\right)^{is}&0\end{bmatrix}ds\,dt\\
&\hspace{32mm}=\int_{0}^{\infty}\int_{0}^{\infty}\left(\frac{4\pi^2\sqrt{\theta(1-\theta)}}{[\cosh(2\pi t)+1][\cosh(2\pi s)+1]}\right)\cos\left(\ln\left(\frac{\phi^{\perp}}{\phi}\right)s\right)\\
&\hspace{40mm}\qquad\times\Big((1-p)\cos\big(\ln(\chi)t\big)+p\cos\big(\ln(\omega)t\big)\Big)ds\,dt\,\sigma_{x},
\end{split}
\]
where 
\[
\chi:=\frac{\theta(1-\phi)}{(1-\theta)\phi}
\qquad\text{ and }\qquad
\omega:=\frac{(1-\theta)(1-\phi)}{\theta\phi}.
\]
At this point, we set $\theta$ and $p$ exactly as in Appendix~\ref{sec:bitflipavgPetz}. Doing so simplifies these expressions to 
\[
\retro^{\JRSWW}_{\alpha,\mF\circ\mE}\left(\begin{bmatrix}0&1\\0&0\end{bmatrix}\right)=\frac{\pi\sqrt{\theta(1-\theta)}}{\sinh(\pi)}\sigma_{x}
\]
and
\[
\left(\retro^{\JRSWW}_{\beta,\mF}\circ\retro^{\JRSWW}_{\alpha,\mE}\right)\left(\begin{bmatrix}0&1\\0&0\end{bmatrix}\right)=\frac{\pi\sqrt{\theta(1-\theta)}}{\sinh(\pi)}\left[\frac{\pi}{2}\left(\frac{4-3\cosh(\pi)-\cosh(3\pi)}{\sinh(\pi)+\sinh(2\pi)-\sinh(3\pi)}\right)\right]\sigma_{x}.
\]
Since the term inside the square parentheses is approximately $1.65$, and therefore not $1$, this shows that  $\retro^{\JRSWW}_{\beta,\mF}\circ\retro^{\JRSWW}_{\alpha,\mE}\ne\retro^{\JRSWW}_{\alpha,\mF\circ\mE}$, i.e., compositionality fails for the JRSWW recovery map. 

Tensoriality also fails for the JRSWW retrodiction family, as we will now illustrate using the same setup as in Appendix~\ref{sec:bitflipavgPetz}, namely $\mE$ is as above, let $(\Alg{A}',\alpha')\xrightarrow{\mE'}(\Alg{B}',\beta')$ be another morphism in $\St$ with $\Alg{B}'=\matr_{2}(\C)=\Alg{A}'$, $\mE'=\Omega_{p'}$, and set  
\[
\alpha':=\begin{bmatrix}\theta'&0\\0&1-\theta'\end{bmatrix}
\quad\text{ and }\quad
\beta':=\begin{bmatrix}\phi'&0\\0&1-\phi'\end{bmatrix}.
\]
Again, set $p=\frac{1}{2}=p'$. Then, by the calculations done in Appendix~\ref{sec:bitflipavgPetz}, 
\[
\begin{split}
\Big(&\retro^{\JRSWW}_{\alpha,\mE}\otimes\retro^{\JRSWW}_{\alpha',\mE'}\Big)\left(\begin{bmatrix}0&1\\0&0\end{bmatrix}\otimes\begin{bmatrix}0&1\\0&0\end{bmatrix}\right)\\
&=\sqrt{\theta\theta^{\perp}\theta'\theta'^{\perp}}\left(\int_{-\infty}^{\infty}\begin{bmatrix}
0&\left(\frac{\theta^{\perp}}{\theta}\right)^{it}\\
\left(\frac{\theta^{\perp}}{\theta}\right)^{-it}&0
\end{bmatrix}d\mu(t)\right)\otimes\left(\int_{-\infty}^{\infty}\begin{bmatrix}
0&\left(\frac{\theta'^{\perp}}{\theta'}\right)^{it'}\\
\left(\frac{\theta'^{\perp}}{\theta'}\right)^{-it'}&0
\end{bmatrix}d\mu(t')\right)\\
&=\sqrt{\theta\theta^{\perp}\theta'\theta'^{\perp}}\left(\int_{0}^{\infty}\frac{2\pi\cos\left(\ln\left(\frac{\theta^{\perp}}{\theta}\right)t\right)}{\cosh(2\pi t)+1}dt\right)\left(\int_{0}^{\infty}\frac{2\pi\cos\left(\ln\left(\frac{\theta'^{\perp}}{\theta'}\right)t'\right)}{\cosh(2\pi t')+1}dt'\right)\sigma_{x}\otimes\sigma_{x}\;,
\end{split}
\]
while
\[
\begin{split}
\retro^{\JRSWW}_{\alpha\otimes\alpha',\mE\otimes\mE'}\left(\begin{bmatrix}0&1\\0&0\end{bmatrix}\otimes\begin{bmatrix}0&1\\0&0\end{bmatrix}\right)
=\sqrt{\theta\theta^{\perp}\theta'\theta'^{\perp}}&\int_{-\infty}^{\infty}
\bigg[\cos\left(\ln\left(\frac{\theta^{\perp}}{\theta}\right)t\right)\cos\left(\ln\left(\frac{\theta'^{\perp}}{\theta'}\right)t\right)\sigma_{x}\otimes\sigma_{x}\\
&+\sin\left(\ln\left(\frac{\theta^{\perp}}{\theta}\right)t\right)\sin\left(\ln\left(\frac{\theta'^{\perp}}{\theta'}\right)t\right)\sigma_{y}\otimes\sigma_{y}\bigg]d\mu(t).
\end{split}
\]
Plugging in the values $\theta=\frac{1}{1+e^{2\pi}}=\theta'$ gives
\[
\Big(\retro^{\JRSWW}_{\alpha,\mE}\otimes\retro^{\JRSWW}_{\alpha',\mE'}\Big)\left(\begin{bmatrix}0&1\\0&0\end{bmatrix}\otimes\begin{bmatrix}0&1\\0&0\end{bmatrix}\right)
=\frac{\pi^2\sqrt{\theta\theta^{\perp}\theta'\theta'^{\perp}}}{\sinh^2(\pi)}\sigma_{x}\otimes\sigma_{x}
\]
and
\[
\begin{split}
\retro^{\JRSWW}_{\alpha\otimes\alpha',\mE\otimes\mE'}\left(\begin{bmatrix}0&1\\0&0\end{bmatrix}\otimes\begin{bmatrix}0&1\\0&0\end{bmatrix}\right)
=\frac{\pi\sqrt{\theta\theta^{\perp}\theta'\theta'^{\perp}}}{2}&\bigg[\left(\frac{1}{\pi}+\frac{1}{\cosh(\pi)\sinh(\pi)}\right)\sigma_{x}\otimes\sigma_{x}\\
&+\left(\frac{1}{\pi}-\frac{1}{\cosh(\pi)\sinh(\pi)}\right)\sigma_{y}\otimes\sigma_{y}\bigg],
\end{split}
\]
which are not equal to each other. This proves that the JRSWW map is not tensorial.

\section{The SS retrodiction family is not stabilizing}
\label{sec:SSnotstabilizing}

In this appendix, we prove Proposition~\ref{prop:SSnotstabilizing}. 
First note that the normalizing and inverting conditions are automatic since they are included in the definition of the Surace--Scandi retrodiction family. 
The remainder of the proof will illustrate that the Surace--Scandi retrodiction family is not stabilizing, specifically not $\circ$-stabilizing. 
In fact, we will not even need to enter the realm of quantum systems to show this. It suffices to work with stochastic matrices. Hence, let $\Alg{A}=\Alg{B}=\Alg{C}=\C^{2}$, let $\alpha,\beta,\gamma$ be the probability vectors given by 
\[
\alpha=\begin{bmatrix}1/2\\1/2\end{bmatrix}
\quad,\quad
\beta=\begin{bmatrix}1/5\\4/5\end{bmatrix}
,\quad\text{ and }\quad
\gamma=\begin{bmatrix}27/50\\23/50\end{bmatrix},
\]
and let $\Alg{A}\xrightarrow{\mF}\Alg{B}$ and $\Alg{B}\xrightarrow{\mF}\Alg{C}$ be represented by stochastic matrices as
\[
\mE=\begin{bmatrix}1/10&3/10\\9/10&7/10\end{bmatrix}
\quad\text{ and }\quad
\mF=\begin{bmatrix}3/10&6/10\\7/10&4/10\end{bmatrix}.
\]
Then $(\Alg{A},\alpha)\xrightarrow{\mE}(\Alg{B},\beta)\xrightarrow{\mF}(\Alg{C},\gamma)$ is a composable pair of morphisms in $\St$, in fact $\CSt$. The Surace--Scandi retrodiction maps in these cases are represented by the following stochastic matrices%
\footnote{For reference, the way these are computed is as follows (it can be done analytically in this case). For concreteness, write $\retro^{\SS}_{\alpha,\mE}$ as a stochastic matrix of the form $\begin{bmatrix}a&b\\1-a&1-b\end{bmatrix}$. The state-preserving condition $\retro^{\SS}_{\alpha,\mE}(\beta)=\alpha$ gives one constraint, which is $\frac{a}{5}+\frac{4b}{5}=\frac{1}{2}$. Detailed balance for $\retro^{\SS}_{\alpha,\mE}\circ\mE$ gives no additional constraints in this case. Requiring that $\retro^{\SS}_{\alpha,\mE}$ is a stochastic matrix and $\retro^{\SS}_{\alpha,\mE}\circ\mE$ has non-negative eigenvalues gives the additional constraint $\frac{1}{2}\le b\le \frac{5}{8}$. Maximizing the determinant of $\retro^{\SS}_{\alpha,\mE}\circ\mE$ with respect to these constraints gives the unique solution $b=\frac{5}{8}$ and $a=0$.}
\[
\retro^{\SS}_{\alpha,\mE}=\begin{bmatrix}0&5/8\\1&3/8\end{bmatrix}
\quad\text{ and }\quad
\retro^{\SS}_{\beta,\mF}=\begin{bmatrix}0&10/23\\1&13/23\end{bmatrix},
\]
while the one associated with the composite $(\Alg{A},\alpha)\xrightarrow{\mF\circ\mE}(\Alg{C},\gamma)$ is 
\[
\retro^{\SS}_{\alpha,\mF\circ\mE}=\begin{bmatrix}25/27&0\\2/27&1\end{bmatrix}.
\]
Since 
\[
\retro^{\SS}_{\alpha,\mE}\circ\retro^{\SS}_{\beta,\mF}=\begin{bmatrix}5/8&65/184\\3/8&119/184\end{bmatrix}\ne\retro^{\SS}_{\alpha,\mF\circ\mE}\;,
\]
compositionality fails. 
In fact, this example also shows that the $\circ$-stabilizing property fails, since every morphism in $\CSt$ is covariant.

\bigskip
\noindent
{\bf Acknowledgements.}
This work is supported by MEXT-JSPS Grant-in-Aid for Transformative Research Areas (A) ``Extreme Universe'', No.\ 21H05183. F.B acknowledges support also from MEXT Quantum Leap Flagship Program (MEXT QLEAP) Grant No.~JPMXS0120319794 and from JSPS KAKENHI, Grants No.~20K03746 and No.~23K03230.
The authors thank James Fullwood, Matteo Scandi, Jacopo Surace, and Karol \.{Z}yczkowski for discussions. The authors also thank two anonymous reviewers for their helpful comments and suggestions.

\addcontentsline{toc}{section}{\numberline{}Bibliography}
\bibliographystyle{quantum}
\bibliography{references-quantum}

\newcommand{\PageBackRef}{}
\PageBackRef
\begin{thebibliography}{10}

\bibitem{vN18}
John von Neumann.
\newblock ``Mathematical foundations of quantum mechanics: {N}ew edition''.
\newblock \href{https://dx.doi.org/10.1515/9781400889921}{Princeton university
  press}. ~(2018). Appearances:~

\bibitem{Lu06}
Gerhart L{\"u}ders.
\newblock ``Concerning the state-change due to the measurement process''.
\newblock \href{https://dx.doi.org/10.1002/andp.20065180904}{Annalen der Physik
  {\bf 15}, 663--670}~(2006). Appearances:~

\bibitem{Kr83}
Karl Kraus.
\newblock ``States, effects, and operations''.
\newblock \href{https://dx.doi.org/10.1007/3-540-12732-1}{Lecture Notes in
  Physics}. Springer Berlin Heidelberg. ~(1983). Appearances:~

\bibitem{Ha75}
S.~W. Hawking.
\newblock ``Particle creation by black holes''.
\newblock \href{https://dx.doi.org/10.1007/BF02345020}{Comm. Math. Phys. {\bf
  43}, 199--220}~(1975). Appearances:~

\bibitem{Ha76}
S.~W. Hawking.
\newblock ``Breakdown of predictability in gravitational collapse''.
\newblock \href{https://dx.doi.org/10.1103/PhysRevD.14.2460}{Phys. Rev. D {\bf
  14}, 2460--2473}~(1976). Appearances:~

\bibitem{Ha82}
S.~W. Hawking.
\newblock ``The unpredictability of quantum gravity''.
\newblock \href{https://dx.doi.org/10.1007/BF01206031}{Comm. Math. Phys. {\bf
  87}, 395--415}~(1982). Appearances:~

\bibitem{Wat55}
Satosi Watanabe.
\newblock ``{Symmetry of Physical Laws. Part III. Prediction and
  Retrodiction}''.
\newblock \href{https://dx.doi.org/10.1103/RevModPhys.27.179}{Rev. Mod. Phys.
  {\bf 27}, 179--186}~(1955). Appearances:~

\bibitem{ABL64}
Yakir Aharonov, Peter~G. Bergmann, and Joel~L. Lebowitz.
\newblock ``Time symmetry in the quantum process of measurement''.
\newblock \href{https://dx.doi.org/10.1103/PhysRev.134.B1410}{Phys. Rev. {\bf
  134}, B1410--B1416}~(1964). Appearances:~

\bibitem{BPJ00}
Stephen~M. {Barnett}, David~T. {Pegg}, and John {Jeffers}.
\newblock ``{Bayes' theorem and quantum retrodiction}''.
\newblock \href{https://dx.doi.org/10.1080/09500340008232431}{J. Mod. Opt {\bf
  47}, 1779--1789}~(2000).
\newblock
  \href{http://arxiv.org/abs/quant-ph/0106139}{arXiv:quant-ph/0106139}.
  Appearances:~

\bibitem{BaKn02}
Howard Barnum and Emanuel Knill.
\newblock ``Reversing quantum dynamics with near-optimal quantum and classical
  fidelity''.
\newblock J. Math. Phys. {\bf 43}, 2097--2106~(2002).
\newblock
  \href{http://arxiv.org/abs/quant-ph/0004088}{arXiv:quant-ph/0004088}.
  Appearances:~

\bibitem{Cr08}
Gavin~E. {Crooks}.
\newblock ``{Quantum operation time reversal}''.
\newblock \href{https://dx.doi.org/10.1103/PhysRevA.77.034101}{Phys. Rev. A.
  {\bf 77}, 034101}~(2008).
\newblock  \href{http://arxiv.org/abs/0706.3749}{arXiv:0706.3749}.
  Appearances:~

\bibitem{APT10}
Yakir Aharonov, Sandu Popescu, and Jeff Tollaksen.
\newblock ``A time-symmetric formulation of quantum mechanics''.
\newblock \href{https://dx.doi.org/10.1063/1.3518209}{Physics Today{\bf
  63}}~(2010). Appearances:~

\bibitem{LeSp13}
Matthew~S. Leifer and Robert~W. Spekkens.
\newblock ``Towards a formulation of quantum theory as a causally neutral
  theory of {B}ayesian inference''.
\newblock \href{https://dx.doi.org/10.1103/PhysRevA.88.052130}{Phys. Rev. A
  {\bf 88}, 052130}~(2013).
\newblock  \href{http://arxiv.org/abs/1107.5849}{arXiv:1107.5849}.
  Appearances:~

\bibitem{Ba14}
Stephen~M. Barnett.
\newblock ``Quantum retrodiction''.
\newblock In Erika Andersson and Patrik {\"O}hberg, editors, Quantum
  Information and Coherence.
\newblock \href{https://dx.doi.org/10.1007/978-3-319-04063-9_1}{Pages 1--30}.
\newblock Springer International Publishing, Cham~(2014). Appearances:~

\bibitem{AZZ15}
Erik {Aurell}, Jakub {Zakrzewski}, and Karol {{\.Z}yczkowski}.
\newblock ``{Time reversals of irreversible quantum maps}''.
\newblock \href{https://dx.doi.org/10.1088/1751-8113/48/38/38FT01}{J. Phys. A:
  Math. Theor. {\bf 48}, 38FT01}~(2015).
\newblock  \href{http://arxiv.org/abs/1505.02259}{arXiv:1505.02259}.
  Appearances:~

\bibitem{OrCe15}
Ognyan Oreshkov and Nicolas~J. Cerf.
\newblock ``Operational formulation of time reversal in quantum theory''.
\newblock \href{https://dx.doi.org/10.1038/nphys3414}{Nature Phys. {\bf 11},
  853--858}~(2015).
\newblock  \href{http://arxiv.org/abs/1507.07745}{arXiv:1507.07745}.
  Appearances:~

\bibitem{OrCe16}
Ognyan Oreshkov and Nicolas~J. Cerf.
\newblock ``Operational quantum theory without predefined time''.
\newblock \href{https://dx.doi.org/10.1088/1367-2630/18/7/073037}{New J. Phys.
  {\bf 18}, 073037}~(2016).
\newblock  \href{http://arxiv.org/abs/1406.3829}{arXiv:1406.3829}.
  Appearances:~

\bibitem{Le16}
Matthew~S. Leifer.
\newblock ``Does time-symmetry in quantum theory imply retrocausality?''.
\newblock  url:~\url{https://pirsa.org/16060059}.
\newblock (accessed:~2022-10-10). Appearances:~

\bibitem{CGS17}
Bob {Coecke}, Stefano {Gogioso}, and John~H. {Selby}.
\newblock ``{The time-reverse of any causal theory is eternal noise}''~(2017).
\newblock  \href{http://arxiv.org/abs/1711.05511}{arXiv:1711.05511}.
  Appearances:~

\bibitem{LePu17}
Matthew~S. Leifer and Matthew~F. Pusey.
\newblock ``Is a time symmetric interpretation of quantum theory possible
  without retrocausality?''.
\newblock \href{https://dx.doi.org/10.1098/rspa.2016.0607}{Proc. R. Soc. A {\bf
  473}, 20160607}~(2017).
\newblock  \href{http://arxiv.org/abs/1607.07871}{arXiv:1607.07871}.
  Appearances:~

\bibitem{Oe19}
Robert Oeckl.
\newblock ``A local and operational framework for the foundations of physics''.
\newblock \href{https://dx.doi.org/10.4310/atmp.2019.v23.n2.a4}{Adv. Theor.
  Math. Phys. {\bf 23}, 437--592}~(2019).
\newblock  \href{http://arxiv.org/abs/1610.09052}{arXiv:1610.09052}.
  Appearances:~

\bibitem{BJP21}
Stephen~M. {Barnett}, John {Jeffers}, and David~T. {Pegg}.
\newblock ``{Quantum Retrodiction: Foundations and Controversies}''.
\newblock \href{https://dx.doi.org/10.3390/sym13040586}{Symmetry {\bf 13},
  586}~(2021).
\newblock  \href{http://arxiv.org/abs/2103.06074}{arXiv:2103.06074}.
  Appearances:~

\bibitem{FSB20}
Dov Fields, Abdelali Sajia, and J\'anos~A. Bergou.
\newblock ``Quantum retrodiction made fully symmetric''~(2020).
\newblock  \href{http://arxiv.org/abs/2006.15692}{arXiv:2006.15692}.
  Appearances:~

\bibitem{DDR21}
Andrea Di~Biagio, Pietro Don{\`{a}}, and Carlo Rovelli.
\newblock ``The arrow of time in operational formulations of quantum theory''.
\newblock \href{https://dx.doi.org/10.22331/q-2021-08-09-520}{{Quantum} {\bf
  5}, 520}~(2021).
\newblock  \href{http://arxiv.org/abs/2010.05734}{arXiv:2010.05734}.
  Appearances:~

\bibitem{CAZ21}
Giulio Chiribella, Erik Aurell, and Karol \.{Z}yczkowski.
\newblock ``Symmetries of quantum evolutions''.
\newblock \href{https://dx.doi.org/10.1103/PhysRevResearch.3.033028}{Phys. Rev.
  Research {\bf 3}, 033028}~(2021).
\newblock  \href{http://arxiv.org/abs/2101.04962}{arXiv:2101.04962}.
  Appearances:~

\bibitem{Ha21}
Lucien {Hardy}.
\newblock ``{Time Symmetry in Operational Theories}''~(2021).
\newblock  \href{http://arxiv.org/abs/2104.00071}{arXiv:2104.00071}.
  Appearances:~

\bibitem{ChLi22}
Giulio Chiribella and Zixuan Liu.
\newblock ``Quantum operations with indefinite time direction''.
\newblock \href{https://dx.doi.org/10.1038/s42005-022-00967-3}{Commun Phys {\bf
  5}, 1--8}~(2022).
\newblock  \href{http://arxiv.org/abs/2012.03859}{arXiv:2012.03859}.
  Appearances:~

\bibitem{SSGC22}
John~H. Selby, Maria~E. Stasinou, Stefano Gogioso, and Bob Coecke.
\newblock ``Time symmetry in quantum theories and beyond''~(2022).
\newblock  \href{http://arxiv.org/abs/2209.07867}{arXiv:2209.07867}.
  Appearances:~

\bibitem{Le07}
Matthew~S. Leifer.
\newblock ``{Conditional Density Operators and the Subjectivity of Quantum
  Operations}''.
\newblock In Guillaume {Adenier}, Chrisopher {Fuchs}, and Andrei~Yu
  {Khrennikov}, editors, Foundations of Probability and Physics - 4.
\newblock \href{https://dx.doi.org/10.1063/1.2713456}{Volume 889 of American
  Institute of Physics Conference Series, pages 172--186}.
\newblock ~(2007).
\newblock
  \href{http://arxiv.org/abs/quant-ph/0611233}{arXiv:quant-ph/0611233}.
  Appearances:~

\bibitem{BuSc21}
Francesco {Buscemi} and Valerio {Scarani}.
\newblock ``{Fluctuation theorems from Bayesian retrodiction}''.
\newblock \href{https://dx.doi.org/10.1103/PhysRevE.103.052111}{Phys. Rev. E
  {\bf 103}, 052111}~(2021).
\newblock  \href{http://arxiv.org/abs/2009.02849}{arXiv:2009.02849}.
  Appearances:~

\bibitem{Ba1763}
Thomas Bayes.
\newblock ``{LII. An essay towards solving a problem in the doctrine of
  chances. By the late Rev. Mr. Bayes, FRS communicated by Mr. Price, in a
  letter to John Canton, A.M.F.R.S}''.
\newblock \href{https://dx.doi.org/10.1098/rstl.1763.0053}{Philos. Trans. R.
  Soc.Pages 370--418}~(1763). Appearances:~

\bibitem{Pearl88}
Judea Pearl.
\newblock ``{Probabilistic Reasoning in Intelligent Systems: Networks of
  Plausible Inference}''.
\newblock \href{https://dx.doi.org/10.1016/C2009-0-27609-4}{Elsevier}. ~(1988).
  Appearances:~

\bibitem{Je90}
Richard~C. Jeffrey.
\newblock ``The logic of decision''.
\newblock University of Chicago Press. ~(1990).
\newblock 2 edition. Appearances:~

\bibitem{HHPBS17}
Dominic {Horsman}, Chris {Heunen}, Matthew~F. {Pusey}, Jonathan {Barrett}, and
  Robert~W. {Spekkens}.
\newblock ``{Can a quantum state over time resemble a quantum state at a single
  time?}''.
\newblock \href{https://dx.doi.org/10.1098/rspa.2017.0395}{Proc. R. Soc. A {\bf
  473}, 20170395}~(2017).
\newblock  \href{http://arxiv.org/abs/1607.03637}{arXiv:1607.03637}.
  Appearances:~

\bibitem{FuPa22}
James Fullwood and Arthur~J. Parzygnat.
\newblock ``On quantum states over time''.
\newblock \href{https://dx.doi.org/10.1098/rspa.2022.0104}{Proc. R. Soc. A {\bf
  478}, 20220104}~(2022).
\newblock  \href{http://arxiv.org/abs/2202.03607}{arXiv:2202.03607}.
  Appearances:~

\bibitem{FuPa22QB}
Arthur~J. Parzygnat and James Fullwood.
\newblock ``From time-reversal symmetry to quantum {B}ayes' rules''~(2022).
\newblock  \href{http://arxiv.org/abs/2212.08088}{arXiv:2212.08088}.
  Appearances:~

\bibitem{EPR}
Albert Einstein, Boris Podolsky, and Nathan Rosen.
\newblock ``Can quantum-mechanical description of physical reality be
  considered complete?''.
\newblock \href{https://dx.doi.org/10.1103/PhysRev.47.777}{Phys. Rev. {\bf 47},
  777--780}~(1935). Appearances:~

\bibitem{Sc35}
Erwin Schr{\"o}dinger.
\newblock ``Discussion of probability relations between separated systems''.
\newblock \href{https://dx.doi.org/10.1017/S0305004100013554}{Math. Proc. Camb.
  Philos. Soc. {\bf 31}, 555--563}~(1935). Appearances:~

\bibitem{Sc36}
Erwin Schr{\"o}dinger.
\newblock ``Probability relations between separated systems''.
\newblock \href{https://dx.doi.org/10.1017/S0305004100019137}{Math. Proc. Camb.
  Philos. Soc. {\bf 32}, 446--452}~(1936). Appearances:~

\bibitem{WJD07}
H.~M. Wiseman, S.~J. Jones, and A.~C. Doherty.
\newblock ``{Steering, Entanglement, Nonlocality, and the
  Einstein-Podolsky-Rosen Paradox}''.
\newblock \href{https://dx.doi.org/10.1103/PhysRevLett.98.140402}{Phys. Rev.
  Lett. {\bf 98}, 140402}~(2007).
\newblock
  \href{http://arxiv.org/abs/quant-ph/0612147}{arXiv:quant-ph/0612147}.
  Appearances:~

\bibitem{LeGa85}
A.~J. Leggett and Anupam Garg.
\newblock ``Quantum mechanics versus macroscopic realism: Is the flux there
  when nobody looks?''.
\newblock \href{https://dx.doi.org/10.1103/PhysRevLett.54.857}{Phys. Rev. Lett.
  {\bf 54}, 857--860}~(1985). Appearances:~

\bibitem{FJV15}
Joseph~F. Fitzsimons, Jonathan~A. Jones, and Vlatko Vedral.
\newblock ``Quantum correlations which imply causation''.
\newblock \href{https://dx.doi.org/10.1038/srep18281}{Sci. Rep. {\bf 5},
  18281}~(2015).
\newblock  \href{http://arxiv.org/abs/1302.2731}{arXiv:1302.2731}.
  Appearances:~

\bibitem{Se07}
Peter Selinger.
\newblock ``Dagger compact closed categories and completely positive maps:
  (extended abstract)''.
\newblock In Proceedings of the 3rd International Workshop on Quantum
  Programming Languages ({QPL} 2005).
\newblock
  \href{https://dx.doi.org/https://doi.org/10.1016/j.entcs.2006.12.018}{Volume
  170, pages 139--163}.
\newblock ~(2007). Appearances:~

\bibitem{CoKi17}
Bob Coecke and Aleks Kissinger.
\newblock ``Picturing quantum processes: A first course in quantum theory and
  diagrammatic reasoning''.
\newblock \href{https://dx.doi.org/10.1017/9781316219317}{Cambridge University
  Press}. ~(2017). Appearances:~

\bibitem{ChJa18}
Kenta Cho and Bart Jacobs.
\newblock ``Disintegration and {B}ayesian inversion via string diagrams''.
\newblock \href{https://dx.doi.org/10.1017/S0960129518000488}{Math. Struct.
  Comp. Sci.Pages 1--34}~(2019).
\newblock  \href{http://arxiv.org/abs/1709.00322}{arXiv:1709.00322}.
  Appearances:~

\bibitem{Fr20}
Tobias Fritz.
\newblock ``A synthetic approach to {M}arkov kernels, conditional independence
  and theorems on sufficient statistics''.
\newblock \href{https://dx.doi.org/10.1016/j.aim.2020.107239}{Adv. Math. {\bf
  370}, 107239}~(2020).
\newblock  \href{http://arxiv.org/abs/1908.07021}{arXiv:1908.07021}.
  Appearances:~

\bibitem{PaBayes}
Arthur~J. Parzygnat.
\newblock ``Inverses, disintegrations, and {B}ayesian inversion in quantum
  {M}arkov categories''~(2020).
\newblock  \href{http://arxiv.org/abs/2001.08375}{arXiv:2001.08375}.
  Appearances:~

\bibitem{Da17}
Tai Danae-Bradley.
\newblock ``{What is a Functor? Definition and Examples, Part 1}''.
\newblock  url:~\url{https://www.math3ma.com/blog/what-is-a-functor-part-1}.
\newblock (accessed:~2022-08-12). Appearances:~

\bibitem{Pe19}
Paolo Perrone.
\newblock ``{Notes on Category Theory with examples from basic
  mathematics}''~(2019).
\newblock  \href{http://arxiv.org/abs/1912.10642}{arXiv:1912.10642}.
  Appearances:~

\bibitem{BaSt11}
John~C. Baez and Mike Stay.
\newblock ``Physics, topology, logic and computation: a {R}osetta {S}tone''.
\newblock In New structures for physics.
\newblock \href{https://dx.doi.org/10.1007/978-3-642-12821-9_2}{Volume 813 of
  Lecture Notes in Phys., pages 95--172}.
\newblock Springer, Heidelberg~(2011).
\newblock  \href{http://arxiv.org/abs/0903.0340}{arXiv:0903.0340}.
  Appearances:~

\bibitem{HeVi19}
Chris Heunen and Jamie Vicary.
\newblock ``Categories for quantum theory: an introduction''.
\newblock \href{https://dx.doi.org/10.1093/oso/9780198739623.001.0001}{Oxford
  University Press}. ~(2019). Appearances:~

\bibitem{Ri17}
Emily Riehl.
\newblock ``Category theory in context''.
\newblock Aurora: Dover Modern Math Originals. Dover Publications. ~(2017).
  Appearances:~

\bibitem{Ma98}
Saunders Mac~Lane.
\newblock ``Categories for the working mathematician''.
\newblock \href{https://dx.doi.org/10.1007/978-1-4757-4721-8}{Volume~5 of
  Graduate Texts in Mathematics, pages xii+314}.
\newblock Springer-Verlag, New York. ~(1998).
\newblock Second edition. Appearances:~

\bibitem{LiWi18}
Ke~Li and Andreas Winter.
\newblock ``{Squashed Entanglement, k-Extendibility, Quantum Markov Chains, and
  Recovery Maps}''.
\newblock \href{https://dx.doi.org/10.1007/s10701-018-0143-6}{Found. Phys {\bf
  48}, 910--924}~(2018).
\newblock  \href{http://arxiv.org/abs/1410.4184}{arXiv:1410.4184}.
  Appearances:~

\bibitem{JRSWW18}
Marius Junge, Renato Renner, David Sutter, Mark~M. Wilde, and Andreas Winter.
\newblock ``Universal recovery maps and approximate sufficiency of quantum
  relative entropy''.
\newblock \href{https://dx.doi.org/10.1007/s00023-018-0716-0}{Ann. Henri
  Poincar\'e {\bf 19}, 2955--2978}~(2018).
\newblock  \href{http://arxiv.org/abs/1509.07127}{arXiv:1509.07127}.
  Appearances:~

\bibitem{Wilde15}
Mark~M. Wilde.
\newblock ``Recoverability in quantum information theory''.
\newblock \href{https://dx.doi.org/10.1098/rspa.2015.0338}{Proc. R. Soc. A {\bf
  471}, 20150338}~(2015).
\newblock  \href{http://arxiv.org/abs/1505.04661}{arXiv:1505.04661}.
  Appearances:~

\bibitem{Pe88}
D{\'e}nes Petz.
\newblock ``Sufficiency of channels over von~{N}eumann algebras''.
\newblock \href{https://dx.doi.org/https://doi.org/10.1093/qmath/39.1.97}{Q. J.
  Math. {\bf 39}, 97--108}~(1988). Appearances:~

\bibitem{Le06}
Matthew~S. Leifer.
\newblock ``Quantum dynamics as an analog of conditional probability''.
\newblock \href{https://dx.doi.org/10.1103/PhysRevA.74.042310}{Phys. Rev. A
  {\bf 74}, 042310}~(2006).
\newblock  \href{http://arxiv.org/abs/0606022}{arXiv:0606022}. Appearances:~

\bibitem{Pe03}
D{\'e}nes {Petz}.
\newblock ``{Monotonicity of Quantum Relative Entropy Revisited}''.
\newblock \href{https://dx.doi.org/10.1142/S0129055X03001576}{Rev. Math. Phys.
  {\bf 15}, 79--91}~(2003).
\newblock
  \href{http://arxiv.org/abs/quant-ph/0209053}{arXiv:quant-ph/0209053}.
  Appearances:~

\bibitem{NgMa10}
Hui~Khoon {Ng} and Prabha {Mandayam}.
\newblock ``{Simple approach to approximate quantum error correction based on
  the transpose channel}''.
\newblock \href{https://dx.doi.org/10.1103/PhysRevA.81.062342}{Phys. Rev. A
  {\bf 81}, 062342}~(2010).
\newblock  \href{http://arxiv.org/abs/0909.0931}{arXiv:0909.0931}.
  Appearances:~

\bibitem{FaRe15}
Omar Fawzi and Renato Renner.
\newblock ``Quantum conditional mutual information and approximate {M}arkov
  chains''.
\newblock \href{https://dx.doi.org/10.1007/s00220-015-2466-x}{Commun. Math.
  Phys. {\bf 340}, 575--611}~(2015).
\newblock  \href{http://arxiv.org/abs/1410.0664}{arXiv:1410.0664}.
  Appearances:~

\bibitem{Je17}
Anna Jen{\v{c}}ov{\'{a}}.
\newblock ``Preservation of a quantum {R}{\'{e}}nyi relative entropy implies
  existence of a recovery map''.
\newblock \href{https://dx.doi.org/10.1088/1751-8121/aa5661}{J. Phys. A: Math.
  Theor. {\bf 50}, 085303}~(2017).
\newblock  \href{http://arxiv.org/abs/1604.02831}{arXiv:1604.02831}.
  Appearances:~

\bibitem{SuSc22}
Jacopo Surace and Matteo Scandi.
\newblock ``State retrieval beyond {B}ayes' retrodiction''.
\newblock \href{https://dx.doi.org/10.22331/q-2023-04-27-990}{{Quantum} {\bf
  7}, 990}~(2023).
\newblock  \href{http://arxiv.org/abs/2201.09899}{arXiv:2201.09899}.
  Appearances:~

\bibitem{Cs91}
Imre Csisz{\'a}r.
\newblock ``Why least squares and maximum entropy? an axiomatic approach to
  inference for linear inverse problems''.
\newblock \href{https://dx.doi.org/10.1214/aos/1176348385}{Ann. Statist. {\bf
  19}, 2032--2066}~(1991). Appearances:~

\bibitem{JRSWW16}
Marius Junge, Renato Renner, David Sutter, Mark~M. Wilde, and Andreas Winter.
\newblock ``Universal recoverability in quantum information''.
\newblock In {2016 IEEE International Symposium on Information Theory (ISIT)}.
\newblock \href{https://dx.doi.org/10.1109/ISIT.2016.7541748}{Pages
  2494--2498}.
\newblock ~(2016). Appearances:~

\bibitem{AEMM19}
Ahmed {Almheiri}, Netta {Engelhardt}, Donald {Marolf}, and Henry {Maxfield}.
\newblock ``{The entropy of bulk quantum fields and the entanglement wedge of
  an evaporating black hole}''.
\newblock \href{https://dx.doi.org/10.1007/JHEP12(2019)063}{J. High Energy
  Phys. {\bf 2019}, 63}~(2019).
\newblock  \href{http://arxiv.org/abs/1905.08762}{arXiv:1905.08762}.
  Appearances:~

\bibitem{CHPSSW19}
Jordan Cotler, Patrick Hayden, Geoffrey Penington, Grant Salton, Brian Swingle,
  and Michael Walter.
\newblock ``Entanglement wedge reconstruction via universal recovery
  channels''.
\newblock \href{https://dx.doi.org/10.1103/PhysRevX.9.031011}{Phys. Rev. X {\bf
  9}, 031011}~(2019). Appearances:~

\bibitem{Pe20}
Geoffrey {Penington}.
\newblock ``{Entanglement wedge reconstruction and the information paradox}''.
\newblock \href{https://dx.doi.org/10.1007/JHEP09(2020)002}{J. High Energy
  Phys.{\bf 2020}}~(2020).
\newblock  \href{http://arxiv.org/abs/1905.08255}{arXiv:1905.08255}.
  Appearances:~

\bibitem{CPS20}
Chi-Fang {Chen}, Geoffrey {Penington}, and Grant {Salton}.
\newblock ``{Entanglement wedge reconstruction using the Petz map}''.
\newblock \href{https://dx.doi.org/10.1007/JHEP01(2020)168}{J. High Energy
  Phys. {\bf 2020}, 168}~(2020).
\newblock  \href{http://arxiv.org/abs/1902.02844}{arXiv:1902.02844}.
  Appearances:~

\bibitem{No21}
Yasunori Nomura.
\newblock ``From the black hole conundrum to the structure of quantum
  gravity''.
\newblock \href{https://dx.doi.org/10.1142/S021773232130007X}{Mod. Phys. Lett.
  A {\bf 36}, 2130007}~(2021).
\newblock  \href{http://arxiv.org/abs/2011.08707}{arXiv:2011.08707}.
  Appearances:~

\bibitem{AkPe22}
Christopher Akers and Geoff Penington.
\newblock ``Quantum minimal surfaces from quantum error correction''.
\newblock \href{https://dx.doi.org/10.21468/scipostphys.12.5.157}{SciPost Phys.
  {\bf 12}, 157}~(2022).
\newblock  \href{http://arxiv.org/abs/2109.14618}{arXiv:2109.14618}.
  Appearances:~

\bibitem{AEHPV22}
Chris Akers, Netta Engelhardt, Daniel Harlow, Geoff Penington, and Shreya
  Vardhan.
\newblock ``The black hole interior from non-isometric codes and
  complexity''~(2022).
\newblock  \href{http://arxiv.org/abs/2207.06536}{arXiv:2207.06536}.
  Appearances:~

\bibitem{Fa01}
Douglas~R. Farenick.
\newblock ``Algebras of linear transformations''.
\newblock \href{https://dx.doi.org/10.1007/978-1-4613-0097-7}{Pages xiv+238}.
\newblock Universitext. Springer-Verlag, New York. ~(2001). Appearances:~

\bibitem{CFS16}
Bob Coecke, Tobias Fritz, and Robert~W. Spekkens.
\newblock ``A mathematical theory of resources''.
\newblock
  \href{https://dx.doi.org/https://doi.org/10.1016/j.ic.2016.02.008}{Inf.
  Comput. {\bf 250}, 59--86}~(2016).
\newblock  \href{http://arxiv.org/abs/1409.5531}{arXiv:1409.5531}.
  Appearances:~

\bibitem{At90}
Michael Atiyah.
\newblock ``The geometry and physics of knots''.
\newblock \href{https://dx.doi.org/10.1017/CBO9780511623868}{Lezioni Lincee}.
  Cambridge University Press. ~(1990). Appearances:~

\bibitem{Fr13}
Daniel Freed.
\newblock ``The cobordism hypothesis''.
\newblock \href{https://dx.doi.org/10.1090/S0273-0979-2012-01393-9}{Bull. Am.
  Math. Soc {\bf 50}, 57--92}~(2013).
\newblock  \href{http://arxiv.org/abs/1210.5100}{arXiv:1210.5100}.
  Appearances:~

\bibitem{Ko14}
Liang Kong.
\newblock ``Anyon condensation and tensor categories''.
\newblock \href{https://dx.doi.org/10.1016/j.nuclphysb.2014.07.003}{Nuclear
  Physics B {\bf 886}, 436--482}~(2014).
\newblock  \href{http://arxiv.org/abs/1307.8244}{arXiv:1307.8244}.
  Appearances:~

\bibitem{BBCW19}
Maissam Barkeshli, Parsa Bonderson, Meng Cheng, and Zhenghan Wang.
\newblock ``Symmetry fractionalization, defects, and gauging of topological
  phases''.
\newblock \href{https://dx.doi.org/10.1103/PhysRevB.100.115147}{Phys. Rev. B
  {\bf 100}, 115147}~(2019).
\newblock  \href{http://arxiv.org/abs/1410.4540}{arXiv:1410.4540}.
  Appearances:~

\bibitem{Ja03}
Edwin~T. Jaynes.
\newblock ``{Probability theory: The logic of science}''.
\newblock \href{https://dx.doi.org/10.1017/CBO9780511790423}{Cambridge
  University Press}. ~(2003). Appearances:~

\bibitem{Ja19}
Bart Jacobs.
\newblock ``The mathematics of changing one's mind, via {J}effrey's or via
  {P}earl's update rule''.
\newblock \href{https://dx.doi.org/10.1613/jair.1.11349}{J. Artificial
  Intelligence Res. {\bf 65}, 783--806}~(2019).
\newblock  \href{http://arxiv.org/abs/1807.05609}{arXiv:1807.05609}.
  Appearances:~

\bibitem{SuToHa16}
David Sutter, Marco Tomamichel, and Aram~W. Harrow.
\newblock ``Strengthened monotonicity of relative entropy via pinched {P}etz
  recovery map''.
\newblock \href{https://dx.doi.org/10.1109/tit.2016.2545680}{IEEE Trans. Inf.
  Theory {\bf 62}, 2907--2913}~(2016). Appearances:~

\bibitem{PaRuBayes}
Arthur~J. Parzygnat and Benjamin~P. Russo.
\newblock ``A non-commutative {B}ayes' theorem''.
\newblock
  \href{https://dx.doi.org/https://doi.org/10.1016/j.laa.2022.02.030}{Linear
  Algebra Its Appl. {\bf 644}, 28--94}~(2022).
\newblock  \href{http://arxiv.org/abs/2005.03886}{arXiv:2005.03886}.
  Appearances:~

\bibitem{GPRR23}
Luca Giorgetti, Arthur~J. Parzygnat, Alessio Ranallo, and Benjamin~P. Russo.
\newblock ``{Bayesian inversion and the Tomita--Takesaki modular group}''.
\newblock \href{https://dx.doi.org/10.1093/qmath/haad014}{Q. J. Math.}~(2023).
\newblock  \href{http://arxiv.org/abs/2112.03129}{arXiv:2112.03129}.
  Appearances:~

\bibitem{Ab15}
Stephen Abbott.
\newblock ``Understanding analysis''.
\newblock \href{https://dx.doi.org/10.1007/978-1-4939-2712-8}{Undergraduate
  Texts in Mathematics}. Springer New York, NY. ~(2015).
\newblock 2 edition. Appearances:~

\bibitem{Ru76}
Walter Rudin.
\newblock ``Principles of mathematical analysis''.
\newblock McGraw-Hill New York. ~(1976).
\newblock 3 edition. Appearances:~

\bibitem{Ma10}
Hans Maassen.
\newblock ``Quantum probability and quantum information theory''.
\newblock In Fabio Benatti, Mark Fannes, Roberto Floreanini, and Dimitri
  Petritis, editors, Quantum Information, Computation and Cryptography: An
  Introductory Survey of Theory, Technology and Experiments.
\newblock \href{https://dx.doi.org/10.1007/978-3-642-11914-9_3}{Pages 65--108}.
\newblock Springer Berlin Heidelberg~(2010). Appearances:~

\bibitem{AwBuSc21}
Clive~Cenxin Aw, Francesco Buscemi, and Valerio Scarani.
\newblock ``Fluctuation theorems with retrodiction rather than reverse
  processes''.
\newblock \href{https://dx.doi.org/10.1116/5.0060893}{{AVS} Quantum Science
  {\bf 3}, 045601}~(2021).
\newblock  \href{http://arxiv.org/abs/2106.08589}{arXiv:2106.08589}.
  Appearances:~

\bibitem{Ba06}
John~C. {Baez}.
\newblock ``Quantum quandaries: A category-theoretic perspective''.
\newblock In Steven French, Dean Rickles, and Juha Saatsi, editors, Structural
  Foundations of Quantum Gravity.
\newblock
  \href{https://dx.doi.org/10.1093/acprof:oso/9780199269693.003.0008}{Pages
  240--265}.
\newblock Oxford U. Press~(2006).
\newblock
  \href{http://arxiv.org/abs/quant-ph/0404040}{arXiv:quant-ph/0404040}.
  Appearances:~

\bibitem{Ka21}
Olav Kallenberg.
\newblock ``Foundations of modern probability''.
\newblock \href{https://dx.doi.org/10.1007/978-3-030-61871-1}{Springer Cham}.
  ~(2021).
\newblock 3 edition. Appearances:~

\bibitem{AbCo04}
Samson Abramsky and Bob Coecke.
\newblock ``A categorical semantics of quantum protocols''.
\newblock In {Proceedings of the 19th Annual IEEE Symposium on Logic in
  Computer Science}.
\newblock \href{https://dx.doi.org/10.1109/LICS.2004.1319636}{Pages 415--425}.
\newblock IEEE~(2004).
\newblock  \href{http://arxiv.org/abs/0402130}{arXiv:0402130}. Appearances:~

\bibitem{OhPe93}
Masanori Ohya and D\'enes Petz.
\newblock ``Quantum entropy and its use''.
\newblock \href{https://dx.doi.org/10.1007/978-3-642-57997-4}{Pages viii+335}.
\newblock Texts and Monographs in Physics. Springer-Verlag, Berlin. ~(1993).
  Appearances:~

\end{thebibliography}

\Addresses

\end{document}